\documentclass[a4paper,11pt]{article}
\pdfoutput=1 
\usepackage{hyperref} 
\usepackage{soul}
\usepackage{url}
\usepackage{xcolor}
\usepackage{pagecolor,lipsum}        % http://ctan.org/pkg/{pagecolor,lipsum}
\usepackage{mdframed}
\usepackage{verbatim}
\usepackage{graphicx}   % subcaption,
\usepackage[T1]{fontenc}             % if needed
\usepackage{mathrsfs}
\usepackage{arydshln}
\usepackage{color}
\usepackage{slashed}
\usepackage{epsfig}
\usepackage{enumerate}
\usepackage{color}
\usepackage{wrapfig}
\usepackage{amsfonts}
\usepackage{multirow}
\usepackage{amsthm}
\usepackage{amsmath}
\usepackage{amssymb}
\usepackage{float}
\usepackage{tikz}
\usepackage[numbers, sort&compress]{natbib}
\usepackage[normalem]{ulem}
\usepackage{multicol}
\usepackage[T1]{fontenc}  % if needed
\usepackage{amsfonts, amsmath, amssymb}
\usepackage{cancel}
\usepackage{enumerate, mathrsfs}
\usepackage{verbatim, hyperref}
\usepackage{graphicx}
\usepackage{color, xcolor}
\usepackage{subfigure}
\usepackage{caption}
\usepackage{multirow, multicol}
\usepackage{lscape}
\usepackage{float}	
\usepackage{pagecolor,lipsum}  % http://ctan.org/pkg/{pagecolor,lipsum}
\usepackage[normalem]{ulem}
\usepackage{epsfig, wrapfig}
\usepackage{mdframed}
\usepackage{textcomp}
\usepackage{arydshln}
\usepackage{lineno}
\hypersetup{
   colorlinks=true,       % false: boxed links; true: colored links
   linkcolor=blue,        % color of internal links
   citecolor=red,         % color of links to bibliography
   filecolor=magenta,      % color of file links
   }

\newcommand{\beq}{\begin{equation}}
\newcommand{\eeq}{\end{equation}}
\newcommand{\bea}{\begin{eqnarray}}
\newcommand{\eea}{\end{eqnarray}}
\newcommand{\nn}{\nonumber}

\begin{document} 
 
%%%%%%%%%%%%%%%%new%%%%%%%%%%%%%%%%%%%%%%%%%%
\begin{center}

{\large \bf Boosted Top Tagging through Flavour-violating interactions at the LHC} \\    
\vskip 0.6cm
Shreecheta Chowdhury$^{a}$%\footnote{shreechetac@srmap.edu.in}
, 
Amit Chakraborty$^{a}$%\footnote{aamit.phy@gmail.com} 
,  
and
Saunak Dutta$^{b}$%\footnote{saunak100@gmail.com}
\vskip 0.6cm

{$^a$  Department of Physics, SRM University-AP, Amaravati, Mangalagiri 522240, India}

\vskip 0.1cm
{$^b$ School of Science and Technology, Vijaybhoomi University, Karjat, Raigad 410201, India.} \\
\end{center}

\vskip 0.3cm

%%%%%%%%%%%%%%%%%%%%%%%%%%%%%%%%%%%%%%55
%\linenumbers 

%\title{\boldmath Probes of Exotic Top Decays in Hadronic Collisions}
%\title{\boldmath Boosted Top Tagging through Flavour-violating interactions at the LHC}

%\author[a,\,\dagger]{Shreecheta Chowdhury,}
%\author[a,\,\dagger\dagger]{Amit Chakraborty,}
%\author[b,c,\,\dagger\dagger\dagger]{and Saunak Dutta,}

%\author[a]{Shreecheta Chowdhury,}
%\author[a]{Amit Chakraborty,}
%\author[b]{and Saunak Dutta,}

%\affiliation[a]{Department of Physics, School of Engineering and Sciences, SRM University AP, Amaravati, Mangalagiri 522240, India}
%\affiliation[b]{International School of Engineering, Lumbini Avenue, Gachibowli, Hyderabad 500032, India}
%\affiliation[b]{ATLAS SkillTech University, Kurla West, Mumbai 400070, India.}

%\emailAdd{${}^{\dagger} shreechetac@srmap.edu.in $}
%\emailAdd{${}^{\dagger\dagger} aamit.phy@gmail.com $}
%\emailAdd{${}^{\dagger\dagger\dagger} saunak100@gmail.com $}

%%%%%%%%%%%%%%%%%%%%%%%%%%%%%%%%%%%%%%%%%%%%%%%%%%%%%%%%%%%%%%%%%%%%%%%%%%%
%=============================================================================================
\begin{abstract}

{
This paper describes a method for detecting a rare top quark decay into a charm quark and a Higgs boson (H), which decays further into b quarks, at the Large Hadron Collider (LHC), and introduces a tagging algorithm to identify boosted tops using large-R jets containing b- and c-tagged elements. We consider the associated production of the top quark with a W-boson and identify different observables to discriminate the signal from the Standard Model (SM) background events. Although our model with improved jet substructure methods outperforms existing approaches to tag such rare decay tops, the improvement in the New Physics reach in terms of $t \to cH$ branching ratio is marginal, even at the high luminosity run of LHC, compared to the existing limits from the LHC 13 TeV data. Although the result seems negative from the point of view of BSM reach, it is significant enough to motivate further studies in the search for the rare top decays at future colliders with higher energy and luminosity. Additionally, the paper utilizes SHAP, a Game Theory-based method, to analyze the contribution of each observable to the classification of events, offering valuable insights into the classifier. The approach presented in this paper is robust in scope and can be effectively implemented for similar probes of physics beyond the SM.
}

%%% 200 word limit 

\end{abstract}

\newpage

\hrule
\tableofcontents
\vskip 1.0cm
\hrule
%\maketitle
%\flushbottom
%%%%%%%%%%%%%%%%%%%%%%%%%%%%%%%%%%%%%%%%%%%%%%55

%=============================================================================================
%%%%%%%%%%%%%%%%%%%%%%%%%%%%%%%%%%%%%%%%%%%%%%%%%%%%%%%%%%%%%%%%%%%%%%%%%%

%\input{introduction}
\section{Introduction}   \label{Sec:Intro}
%%%%%%%%%%%%%%%%%%%%%%%%%%%%%%%%%%%%%%%%%%%%%%%%%%%%%%%%%%%%%%%%%%%%%%%%%%%%%%%%%%%%%%%
The Standard Model (SM) of Particle Physics is, by far, the most successful formalism to unify and address three of the four known fundamental interactions in nature. In its framework, the Flavor-Changing Neutral Current (FCNC) refers to the exchange of a boson of electromagnetic charge zero in the interactions that lead to a change of the flavor of fermions initially taking part in it. Since in the formulation of the SM, the mass and gauge eigenstates of the fermions for the interactions involving FCNC are identical, the works of Glashow, Iliopoulos and Maiani \cite{Glashow:1970gm} established that there cannot be any interaction involving FCNC in SM at tree level. 
%However, since the mass and gauge eigenstates corresponding to the $SU(2)$ group are different, the effect of the flavor-changing charge current (FCCC) predominantly occurs in the quark sector within the SM \cite{ALTARELLI:2005zv}. The probability of the transition of a down-type quark of a given family to that of any family following an FCCC-interaction is given by the CKM matrix, named after the ones who proposed it, Cabibbo, Kobayashi and Maskawa \cite{PhysRevLett.10.531, 10.1143/PTP.49.652}.
Hence, the quest for the signatures of the processes that enable FCNC is an effective probe of new physics.

%\textcolor{blue}{The quest on the signatures of the processes which enable Flavour Changing Neutral Current(FCNC) are effective probes of New Physics. Since in the formulation of the SM, the mass and gauge eigen-states of the fermions for the interactions involving FCNC are identical, the works of Glashow, Iliopoulos and Maiani \cite{Glashow:1970gm} established that there cannot be any interaction involving FCNC in SM at tree level. However, according to Cabibbo, Kobayashi and Maskawa \cite{PhysRevLett.10.531, 10.1143/PTP.49.652} transition among different flavours of quarks are possible through charge current (Flavour Changing Charge Current (FCCC)) and the probabilities are given by the elements of CKM matrix.
%Ref.\cite{Kiers:2011sv, Bardhan:2016txk, Chen:2023eof, Altmannshofer:2019ogm, Hung:2017tts, Larios:2006pb, Banerjee:2018fsx, Kim:2015zla} (and the references therein) showcase a comprehensive analysis in this direction involving the decay of the top quark.
%With the discovery of the top quark at Tevatron in 1995 \cite{topdiscov1, topdiscov2}, the observations of all the members of the three fermionic families carrying electromagnetic charges were completed. 

The top quark is the heaviest of all particles accommodated in the SM framework, with a mass of around 172 GeV \cite{article:tawj} and an electromagnetic charge of $Q=2/3$ units. It also has a very short lifetime ($\sim 5 \times 10^{-25}$ sec) \cite{Quadt2007TopQP}, even shorter than the time required for hadronization ($\sim 10^{-24}$ sec) \cite{aubert2012particle}. Therefore, a top quark produced in a particle collider fails to hadronize to form bound states like other quarks, and rather decays as a free quark. Its decay therefore provides an opportunity to study new physics signatures for a quark in its bare state, unaffected by hadronization. In SM, the top quark decays to a bottom quark and a W boson, $t \to b W^{+}$ or $\bar{t} \to \bar{b} W^{-}$ 
with a branching ratio of ~100\%. The branching fractions for the other modes of top decay are $\mathcal{O}(0.1)\%$ or less.  
%$t \to s W^{+}$ (or ${\bar{t} \to \bar{s} W^-}$) and $t \to d W^{+}$ (or ${\bar{t} \to \bar{d} W^-}$) are respectively $\sim 0.1\%$ and $\sim 0.01\%$ \cite{CMS:2014mxl}. 
%The W boson thus produced may undergo either a leptonic decay $(W^{+} \to \ell^{+} \nu_{\ell})$ or a hadronic decay $(W^{+} \to q q^{\prime})$, where $q$ and $q^{\prime}$ are quarks of the up- and down-type belonging to the same weak isospin doublet. 
The decay of a top quark to an up-type quark (of the same charge) through interaction involving FCNC is highly suppressed \cite{fcncBr1, fcncBr2, fcncBr3}. Despite the loop suppression of FCNC decays in the SM, there are certain Beyond the Standard Model (BSM) theories which lead to somewhat enhanced branching of such exotic decay channels of top, and hence a precise measurement of such decays would either favor or rule out those theories. We refer to \cite{Kiers:2011sv, Bardhan:2016txk, Chen:2023eof, Altmannshofer:2019ogm, Hung:2017tts, Larios:2006pb, Banerjee:2018fsx, Kim:2015zla} (and the references therein) for a comprehensive analysis in this direction. In recent studies of the 13 TeV data by the ATLAS and CMS collaborations, the current limits on the $t-c-H$ coupling through single top quark production can be found in \cite{ATLAS:2022gzn,CMS:2021hug,CMS:2021gfa,CMS:2017bhz}. 

In this paper, we perform a model-independent analysis of the signature involving $t \to c H$ decay, a potential probe for New Physics. The coupling between the top and charm with the Higgs boson in principle can contribute to different production modes. However, in this work, we consider the contribution of this coupling to FCNC-mediated top decay. The role of $t-c-H$ coupling in the production of the top quarks, preserving its SM decay modes, was explored in \cite{Greljo:2014dka}. Our work presents a complementary approach to the previously mentioned article. Therefore, the scope of our study does not constrain the $t-c-H$ coupling, nor is it an exhaustive exploration of the production channels to which the said coupling may lead. We rather restrict and redirect our analysis towards efficient identification of the $t \to c H$ decay mode (in the boosted regime) as a trivial consequence of the $t-c-H$ coupling.{Specifically, we ask the following questions : Are the existing jet-substructure-based toptagging algorithms efficient enough to probing the FCNC decays of the top quark? What modifications can be incorporated into the tagger to improve the new physics reach?} 
%
%The role of $t-c-H$ coupling in the production of the top quarks, preserving its SM decay modes, was explored in \cite{Greljo:2014dka}. Our work presents a complementary approach to the previously mentioned article.

We consider the production of a top quark accompanied with a W boson, $pp \to t W $, at the Run-$\mathrm{II}$ of CERN Large Hadron Collider (LHC) with a collision energy of 13 TeV, with the $W$ boson undergoing muonic decay ($W^- \to \mu^- \bar{\nu}_\mu$). We focus on the phase-space region where the produced top quark possesses a high transverse momentum ($viz$. $p_{T} > $ 200 GeV). As a consequence, highly boosted muons and a large missing transverse energy can be set as a trigger for such events. The boosted top quark would eventually decay to a c-quark and Higgs boson, which will be highly collimated within a large radius (R) along the top quark's direction of motion. The Higgs boson thus produced would predominantly decay to $b\bar{b}$ pair. Thus, in ideal scenario, one would expect to find three subjets, one of which is c-tagged and the rest b-tagged. The dominant background process of SM (among all other potential ones) would mainly consist of top quarks decaying to $bW$, where the $W$ boson so produced would undergo a decay to $\bar{c} s$ \cite{CMS:Wcs}. For such events, too, one would ideally obtain three subjets within the large-R jet: one c-tagged, one b-tagged, and the rest a light jet. Considering the mistagging of the light jet as a b-tagged jet, the subjet structure of the background might resemble that of the signal. Note that the invariant mass of the two b-tagged subjets for the signal would peak at the Higgs boson mass (125 GeV), for the b-quarks originated from Higgs boson decay. In contrast, such a peak is not expected to appear for the SM background(s). This contrast of signal and background events would be a potential discriminator. For completeness, we have also considered QCD multijet backgrounds, $pp \to jj$. In parallel, we also simulate $pp \to Wg$ events, $W$ undergoing muonic decay. For both of these backgrounds, jets with the largest transverse momentum are considered as top-candidate jets, and their individual contributions as the background are weighted by their respective production cross sections. A multivariate analysis (MVA) using the Boosted Decision Tree (BDT) algorithm, for events with at least 1 b-tagged and 1 c-tagged subjets constituting the boosted large-R top jet, yields a substantial signal efficiency of 80\% over a significant background rejection of 90\%.

{
As mentioned, our work focuses on proposing a boosted top tagger, sensitive to the FCNC decay of a top quark, $t \to cH$. The following observations are made based on our analysis: 
\begin{itemize} 
\item Existing cut-based boosted taggers (for hadronically decaying tops) aim to reconstruct the W mass and subsequently the top mass using substructure-based observables \cite{Chakraborty:2023dhw, JHToptagger, Plehn:2011tg, Thaler:2008jutagger, Kasieczka:2017nvn, Butter:2017cot, Macaluso:2018tck, Kasieczka:2019dbj, Dreyer:2020brq, Andrews:2021ejw, Chakraborty:2020yfc, Alvarez:2022qoz, Bhattacherjee:2022gjq}. They have been successfully implemented in different experimental studies \cite{ATLAS:2021wkg, ATLAS:2020szu, ATLAS:2020lks, ATLAS:2018wis, CMS:2017ucf, CMS:2021beq}.
\item Several studies have been conducted that propose different machine learning (ML)-driven top-candidate jet tagging algorithms \cite{Kasieczka:2019dbj, Sahu:2023uwb, Butter:2017cot, Kasieczka:2017nvn}. They are principally aimed at the identification and reconstruction of top-candidate jets from respective hadronic decays in compliance with SM. 
\item  We find that existing top-tagging methods (both cut-based and ML-based) are not sensitive enough to tag a boosted top candidate jet when the top decays through the $cH$ mode, and therefore there is a need to modify existing taggers to investigate the FCNC decay. 
\end{itemize} 
The tagger we propose in this work distinguishes the top-candidate jets that undergo suppressed FCNC decays, from the ones that decay to $b W$. By comparing the performance of the proposed tagger with existing ML-based algorithms using the ROC (Receiver Operating Characteristic) curve, we observe that top taggers built using high-level objects (viz. observables constructed using the information of the reconstructed jet and its substructure, and applying flavor tagging at the subjet level) are more efficient for probing the $cH$ mode compared to the ones using low-level information like Jet image or GNN based setups. The performance of the proposed tagger has also been tested considering both the irreducible background consisting of $t \to bW$ samples and the QCD background, which has a very high production cross section. 
Although the proposed method improves sensitivity for detecting FCNC-mediated top quark decays using jet-substructure techniques, it results in a lower signal significance in terms of the limits on $t \to cH$ branching ratio, compared to the current limits from the resolved analyses by the ATLAS and CMS collaborations at the LHC. While it is insufficient to confirm any evidence or make a discovery of new physics, it is significant enough to motivate future collider searches for rare FCNC-mediated top decays.} Finally, we introduce SHAP (SHapley Additive exPlanations), a formalism developed from the Game Theory, that provides an insight into the contribution of different discriminators in our tagger, thereby unfolding the respective sets of features sensitive to the classification.

%\textcolor{red}{How do we understand the decision of the model on the classification task?} To address this, we introduce SHAP (SHapley Additive exPlanations), a formalism developed from the Game Theory, that provides an insight into the contribution of different discriminators in our tagger, thereby unfolding the respective sets of features sensitive to the classification.

This paper progresses as follows: Section \ref{Sec:Mod} presents the effective coupling between the top, charm, and the Higgs boson in a model-independent manner. Section \ref{Sec:EvSim} discusses the prescriptions followed for the event generations and the selection of Boosted jets. The results are presented in Section \ref{Sec:Res}, and the possible interpretation of the trained machine learning model using Shapley is discussed in Section \ref{Sec:Shap}. The paper ends with the Section \ref{Sec:Conc}, presenting a summary of the principle, techniques and effectiveness of the proposed tagging algorithm along with the potential areas thereby unveiled to explore.

%Here we follow a similar strategy: construct several substructure-based observables of the top candidate jets, apply b- and c-tagging methods, aim to reconstruct the Higgs boson using the b-tagged subjets, and finally combine two b-tagged subjets with the c-tagged subjet to reconstruct the top quark. 

%\sout{As an alternative implementation of our tagger, we test its efficacy in BSM probes involving similar topolgy: we consider the vector-like quark $Y$ of electromagnetic charge $-\frac{4}{3}$ decaying to a $b$-quark and a $W$-boson, $W^-$ decays further to $c\bar{s}$. Therefore, for a heavy $Y$ ($m_Y$ = 600 GeV satisfying the current bound \cite{CMS:2018dcw}), the large-R jet constructed using the decay products of $Y$, will include both b-jets and c-jets as substructures. Following similar strategy, our proposed tagger reconstructs the $W$-boson using the c- and light jets, and subsequently the $Y$-boson which includes the b-, c- and light jets.} 

%=============================================================================================
\section{The Model}    \label{Sec:Mod}
We focus on the signature of an exotic decay channel of the top quark involving Flavour Changing Neutral Current (FCNC) in a model-independent way. Similar probes of New Physics, through electroweak and effective couplings involving FCNC-mediated top decay, have been extensively performed in \cite{Kao:2011aa,Buchkremer:2013bha,Degrande:2011rt,Greljo:2014dka,Bardhan:2016txk}. The effective coupling for $t \to c H$ can be parameterized as 
\beq \label{Eqn:Lgrng}
\mathcal{L}_{tcH} = - \eta_L \bar{t}_L c_R H + \eta_R \bar{c}_L t_R H  + \mathit{h.c.}
\eeq
where the coupling strengths, $\eta_L$ and $\eta_R$ are associated respectively with the left and right chiral top quarks. We emphasize that the above Lagrangian (Eqn. \ref{Eqn:Lgrng}) can be a part of a gauge-invariant UV-complete theory, which takes the form after spontaneous breaking of the electroweak symmetry. The coupling parameters, $\eta_L$ and $\eta_R$, provide a non-standard contribution towards the total decay width of the top quark, and therefore, precise measurement of the decay width of top along with direct search for the $t \to c H$ mode put severe constraint on these couplings, as explored in \cite{Greljo:2014dka}. In this paper, we choose to probe this coupling through the top decay in a phase space region where it is sufficiently boosted, leading to our proposed tagger sensitive to the FCNC-mediated top decay, enabling the BSM probes manifested in large-R jets with similar jet substructures. We follow the prescription of the effective model available in FeynRules \cite{feynrules1, feynrules2} with the $t-c-H$ coupling parameterized by the parameter $\kappa_{tcH}$. The variation in $\kappa_{tcH}$ leads to different branching fractions of the $t \to cH$ decay.
      
%%%%%%%%%%%%%%%%%%%%%%%%%%%%%%%%%%%%%%%%%%%%%%%%%%%%%

%=============================================================================================
%\input{EventGeneration}

\section{Event Generation and Selection of Boosted Jets}  \label{Sec:EvSim}

\begin{figure}[!htb]
\centering
\includegraphics[width=0.9\textwidth]{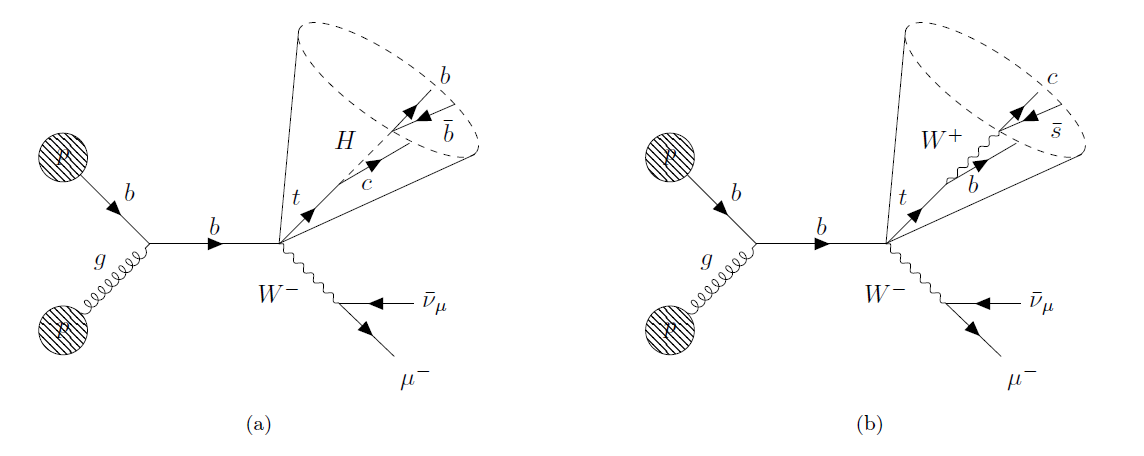}
\caption{\label{Fig:1}Tree-level Feynman diagrams corresponding to the s-channel production of signal ($t\to c H$) and most dominant irreducible background ($t\to b W$) events.}
\end{figure}

We simulate the events where the top quark is produced in association with a W boson at the LHC. These events are used for both the signal and the dominant background. As shown in Figure \ref{Fig:1}, both the signal and the dominant irreducible SM background share the identical tree-level Feynman diagram for the production of the top and the associated $W$. The difference lies in the decay of the generated top: for the signal the top quark decays to a c-quark and a Higgs boson (H) which subsequently decays to the $b\bar{b}$ pair, whereas for the background the top undergoes a decay to a b-quark and $W$ boson with the $W$ further decaying to c and s-quarks. In principle, $W$ can decay to all possible hadronic modes. However, for this work, we considered the $W$-decay to $c-s$ mode, since our proposed top tagger requires a substructure of a c-tagged jet inside the boosted large-R jet. We emphasize here that the cross section obtained for the event generation, for both the signal and background processes, is leading order(LO) in the strong coupling constant $\alpha_s$.

%%%%%%%%%%%%%%%%%%%%%%%%%%%%%%%%%%%%%%%%%%%%%%%%%%%%%%

%%%%%%%%%%%%%%%%%%%%%%%%%%%%%%%%%%%%%%%%%%%%%%%%%%%%%%
The cascade of sequential decays that follow from the production of the top quark in association to a $W$ boson for the signal and dominant SM background events are enlisted in Table \ref{tab:simulation}. Note that the production of $W$ bosons in association with a light jet ($Wj$) and QCD dijet events also contributes to the list of background events, which we have also simulated with an appropriate choice of generation-level cuts on the hard processes. The total contribution of the SM backgrounds is obtained by assigning proper weight to these processes according to the respective production cross sections. 
Further details on the simulation and reconstruction procedure of events are provided in the Appendix \ref{Sec:App1}. 

%---------------------------------------
\begin{table}[!htb]
\centering
\begin{tabular}{ |c|c|c|} 
 \hline
 Sample type& Process simulated & Generation level cuts \\ 
 \hline 
 \hline 
Signal & $p p \to t W^-, ~t \to c H, ~W^- \to \mu^- \bar{\nu}_\mu, H \to b \bar{b}$ & $p^{top}_{T, min}$ = 350 GeV \\ 
%       & $p p \to \bar{t} W^+, ~\bar{t} \to \bar{c} h, ~W^+ \to \mu^+ \nu_\mu, h \to b \bar{b}$ & \\ 
 \hline 
Background 1  & $p p \to t W^-, ~t \to b W^+, ~W^- \to \mu^- \bar{\nu}_\mu, W^+ \to c \bar{s}$ & $p^{top}_{T, min}$ = 350 GeV \\ 
%	& $p p \to \bar{t} W^+, ~\bar{t} \to \bar{b} W^-, ~W^+ \to \mu^+ \nu_\mu, W^- \to \bar{c} s$ & \\
%	\cline{2-3} 
 \hline 
Background 2 & $p p \to j j$ & $p^{j}_{T, min}$ = 350 GeV \\
%	\cline{2-3} 
 \hline 
Background 3 & $p p \to j j$ & $p^{j}_{T, min}$ = 500 GeV \\ 
%	\cline{2-3} 
 \hline 
Background 4 & $p p \to g W^{\pm}, ~W^{\pm} \to \mu^{\pm} \nu_\mu$ & $p^{gluon}_{T, min}$ = 350 GeV \\
 \hline 
\end{tabular}
\caption{Details of the signal and background processes simulated for our study. Note that, although this table and Figure \ref{Fig:1} indicates the processes involving one type of particles, we have simulated events with charge conjugated states as well.}
\label{tab:simulation}
\end{table}

%========================================

%=============================================================================================

%\section{Results and Discussion} \label{Sec:Res} 
\section{Collider Analysis and Discussion} \label{Sec:Res} 

The previous section was consecrated to the discussion of generation of hard-scattering events for both the signal and SM backgrounds, followed by the detector simulations, along with the specifications of cuts and triggers imposed for the selection of the events. In this section, we present the results obtained from our analysis after the proper imposition of cuts. This section is redistributed into several subsections, each categorizing a particular stage of the analysis.

\subsection{Analysis of Parton Level Events}  \label{SubSec:HrdPro}

We commence our discussion with the presentation of distributions of different observables for parton-level scattering, which motivates the choice of cuts we have imposed in the subsequent stages of our analysis. We take into account the context of kinematics for both the signal (i.e. $p p \to t W^-, ~t \to c H$) and the dominant SM background (i.e., $p p \to t W^-, ~t \to b W^+$) for the starter. Exploring different observables at the parton level, we observe the following.

\begin{itemize}
\item The generation level cut on the minimum $p_T$ of top quark at 350 GeV generates the signal and background events in the desired region of phase space quite efficiently. The muon originating from the W-decay also found out to be quite boosted, and therefore, can be used to trigger these events along with large missing transverse energy.    

\item The heavier counterpart of the top decay in both signal and background events, namely the Higgs boson and the W-boson respectively, follow similar behaviour in the boosted regime with $p_T > 300$~GeV. We however also observe slightly higher proportion of boosted Higgs boson than the W-boson, the reason for which can be attributed to the higher disparity in mass of the top decay products. 

\item 
The b and c-quarks have different origins for the signal and dominant SM background: b-quarks appear from the Higgs boson decay for the signal and the boosted top-decay for the background, whereas the c-quark appears from the boosted top-decay for the signal and $W$-boson decay for the background. In Figure \ref{Fig:5}, we show the parton-level $p_T$ distribution of the b and c-quarks produced in both the signal and background events. For the signal events, this c-quark is produced in association with the massive Higgs boson from the top-decay, has a shorter tail (till $\sim$400 GeV) compared to its counterpart in the background events which are produced from the $W$-decay (see the left panel). The reason behind this can be attributed to the higher disparity in mass for the top decay products in signal and background events. It is found that the b-quark for the signal and the background behaves almost in a similar manner (see the right panel of Figure \ref{Fig:5}).

\begin{figure}[!htb]
\centering
\includegraphics[width=.4\textwidth]{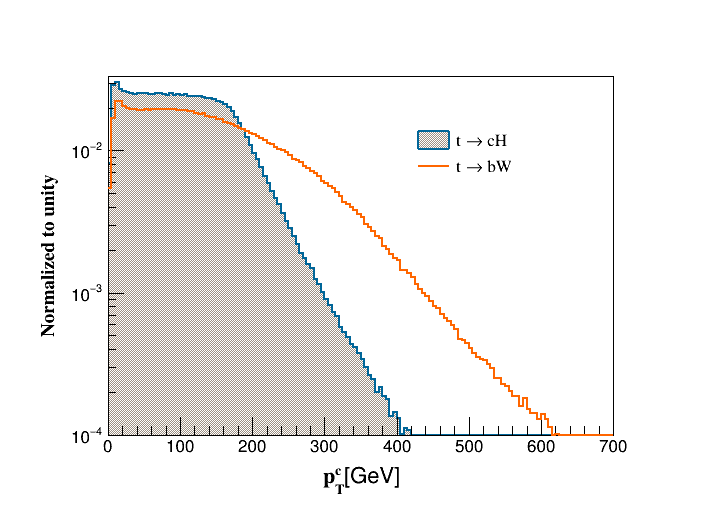}
\includegraphics[width=.4\textwidth]{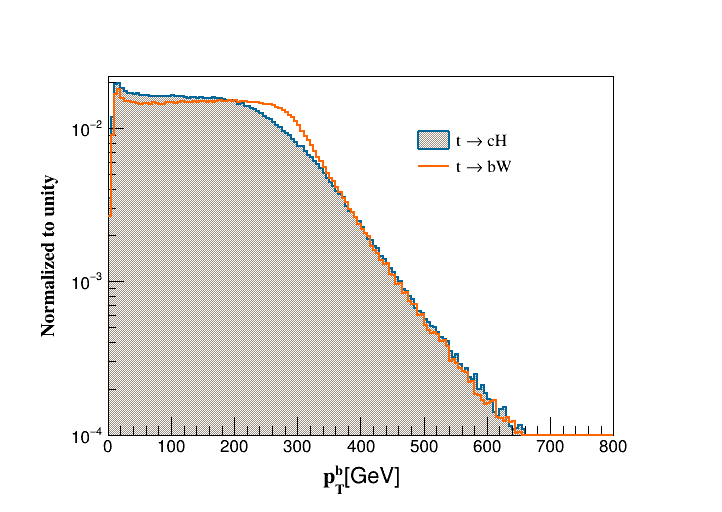}
\caption{\label{Fig:5} Normalized $p_T$ Distribution for the c and b-quarks, at parton-level, for the Signal and the Background events simultaneously. \textit{Left}: $p_{T}$ Distribution of the c-quark originated from the top-decay for the signal and $W$-decay for the background. \textit{Right}: $p_{T}$ Distribution of the leading b-quark originated from the Higgs-decay for the signal and top-decay for the background.}
\end{figure}

\item The choice of jet radius is very important for the subsequent analysis. Therefore, we study the angular separation between the top decay products, namely $\Delta R$ between the charm and Higgs boson for the signal and between the bottom quark and the W-boson for the dominant background. As Figure \ref{Fig:6} suggests, the $\Delta R$ distribution peaks around 0.4 and 0.7 for the signal and background events respectively. We therefore, cluster the events with jet radius R = 1, and perform the detailed collider analysis.

\begin{figure}[!htb]
\centering
\includegraphics[width=.4\textwidth]{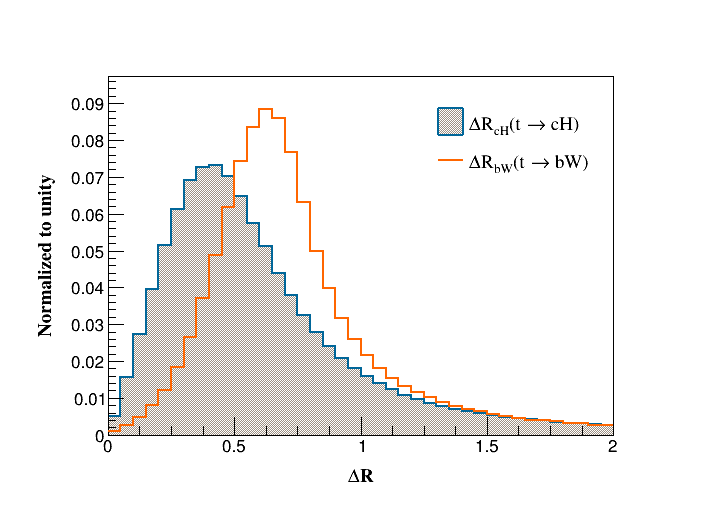}
\includegraphics[width=.4\textwidth]{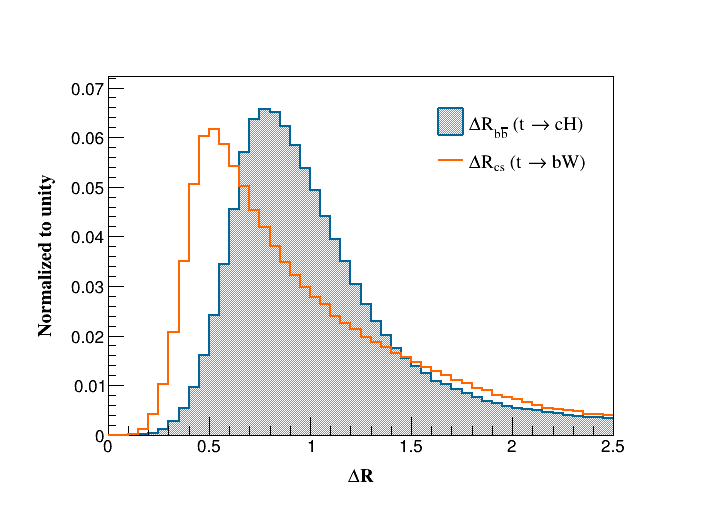}
\caption{Left: The $\Delta R$ distribution between the decay products of the top-quark for the signal and the background respectively. Right: The $\Delta R$ distribution between the decay products of the Higgs boson and $W$-boson for the signal and background events respectively.}  \label{Fig:6}
\end{figure}
\end{itemize}

On a passing note, we also consider other possible backgrounds, for example QCD dijet events and $Wg$ production with $W$ decaying leptonically. Depending on the $p_T$ of the QCD jets (quarks and gluons at the parton level), after showering and hadronization, they can also exhibit a three-prong structure which will mimic the signal topology after light jets are mistagged as b- and c-tagged jets. So, from the next section onward, we take into account all possible SM processes that can be considered as potential SM backgrounds.

%%%%%%%%%%%%%%%%%%%%%%%%%%%%%%%%%%%%%%%%%%%%%%%%%%%%%%%%%
\subsection{Detector Level Analysis} \label{SubSec:JMult}

The observations presented in Section \ref{SubSec:HrdPro} provided convincing evidence to infer that the jets formed from the decay of the highly boosted top quark are collimated and collectively form a large-R jet with a three-prong structure in an ideal case. Hence, starting with the stable decay products for each event as obtained after the hadronization and parton showering, we obtain reconstructed objects taking into account detector effects. We now outline the input variables (or features) used to analyze the background and signal events. 

\begin{itemize}
\item{\bf {Large-R jets:}}

The jets are obtained after clustering EFlow objects using the anti-kT jet algorithm\footnote{We have also implemented Cambridge-Achen (C/A) jet algorithm \cite{Dokshitzer:1997in}, in parallel to anti-kT for jet reconstruction, and the observations do not show significant deviation between the two approaches.} with the radius parameter R = 1.0. As discussed before, our aim is to detect three-prong jet substructures of a large-R jet of mass in the vicinity of the top mass. Once the events are reconstructed to large-R jets, we apply the following selection criteria to analyze the events, 
\begin{itemize} 
\item Select the reconstructed large-R jet with $p_T >$ 400 GeV  having mass in the range of [140 - 200] GeV; it's the candidate topjet. 
\item Recluster the constituents of the candidate jet within radius $ R_{subjet} \sim 0.3$ with $p_{T,~min}^{subjet} = 20~\text{GeV}$, and apply b- and c-tagging algorithms at the subjet level. 

\item Select the leading large-R jet that contains at least one b- and one c-tagged subjets.
\end{itemize} 

%So we select jets with minimum transverse momentum of $p_{T,~min}^{jet} = 400~\text{GeV}$ having mass in the range of [140 - 210] GeV, and reclustered the constituents of the large-R jet to obtain subjets of radius $ R_{subjet} \sim 0.3$ with $p_{T,~min}^{subjet} = 20~\text{GeV}$. Among these subjets, in an ideal case, two b- and one c-tagged subjets should be present inside the large-R jet for the signal events. However, we take a conservative approach and select those large-R jets which have at least one b- and c-tagged subjet inside it. 

Analyzing the subjet kinematics, we also observe that the transverse momentum of the leading subjets tagged with the b and c jet ($p^{b-jet}_T$ and $p^{c-jet}_T$) and the number of light jets ($N_{light-jet})$ can be used to discriminate the signal events from the background ones. Therefore, these observables are also used for subsequent analysis. The details of kinematics of the jets and their flavor tagging are described in the appendix \ref{Sec:App2}. 

\item{\bf {N-Subjettiness:}}

We carry forward our analysis with the observable N-Subjettiness \cite{NSubjettiness} of a given boosted jet, with N denoting the number of subjet axes considered within the large-R jet. It provides a measure of the aggregate of the angular distances of the constituents of the jet from their respective nearest subjet axis. It is defined as \cite{NSubjettiness},
\beq \label{Eqn:N-SubJness}
\tau _{N}(\beta) ~= ~ \frac{1}{R_{J}^{\beta} \sum_{k} p_{T,k}} ~\sum_{k} \left( p_{T,k} \times min \{ \Delta R_{1,k}^{\beta}, ~\Delta R_{2,k}^{\beta}, ~..., ~\Delta R_{N,k}^{\beta} \} \right),
\eeq
where k runs over the constituents of the given large-R jet, $p_{T,k}$ being the corresponding transverse momentum, $\beta$ is an angular weighting exponent, $R_{J}$ is the large-R radius, and $\Delta R_{j,k} ~=~ \sqrt{ (\Delta y)_{j,k}^2 + (\Delta \phi)_{j,k}^2 }$ ($y:$ pseudo-rapidity and $\phi:$ azimuthal angle of a concerned particle/axis) is the angular separation between the $k^{th}$ constituent and the $j^{th}$ subjet axis. In order to estimate how distinct the three subjet axes for the large-R jet are, we obtain the distribution for the ratio of $\frac{\tau_{N}}{\tau_{N-1}}$ denoted $\tau_{N (N-1)}$ for $N ~=~ 2, ~3$ with $\beta ~=~ 1$. Figure \ref{Fig:14} (left) sketches the distribution of $\tau_{21}$, which gives an estimate of the proximity of the large R jet to a two-prong structure with respect to a single-prong substructure. The signal and the irreducible background gain a distinct peak at $\sim$0.5 while the QCD events show a left-skewed distribution with flatter peaks towards the right. The behavior of $\tau_{32}$, which provides an estimate of the 3-prong nature of the large-R jet over the 2-prong trait, in Figure \ref{Fig:14} (right) shows a behavior similar to that observed for $\tau_{21}$. The sharp peak for the QCD events towards the right suggests that these jets have essentially 2-prong structures. 
\begin{figure}[!htb]
\centering   % \begin{center}/ \end{center} takes some additional vertical space
\mbox{
{\includegraphics[width=.4\textwidth]{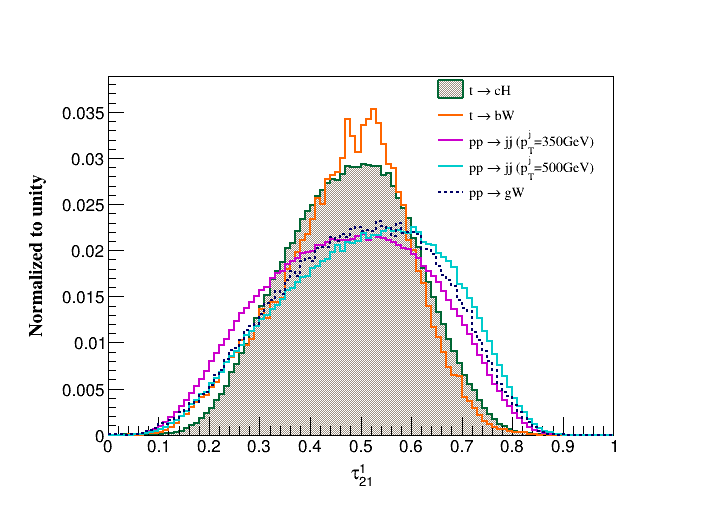}  }
{\includegraphics[width=.4\textwidth]{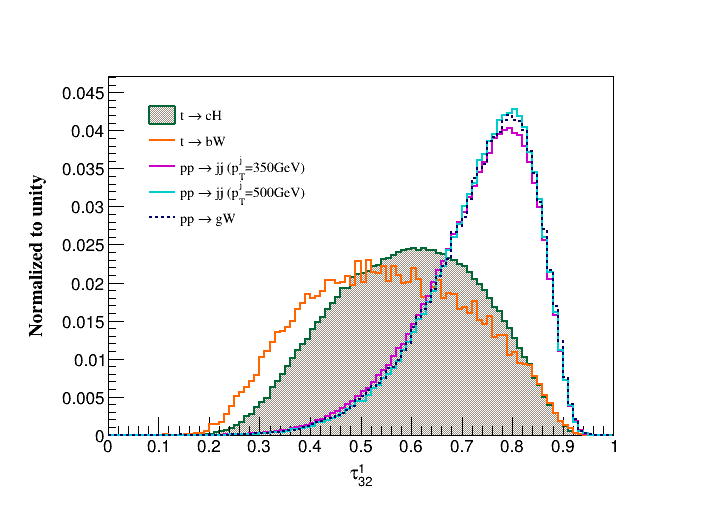}  }}
\caption{\label{Fig:14}\textit{Left}: Normalized distribution for $\tau_{21} ~=~ \frac{\tau_{2}}{\tau_{1}}$, \textit{Right}: Normalized distribution for $\tau_{32} ~=~ \frac{\tau_{3}}{\tau_{2}}$, with the angular weightage $\beta ~=~ 1$, for signal, irreducible background and QCD events.}
\end{figure}

%----------------------------------------------------------
%----------------------------------------------------------

%---------------------------------------------------------------------------
\item{\bf Energy fraction of the subjets:}

Next, we calculate the fraction of the large-R jet energy repatriated among its constituent subjets. The energy fraction possessed by a subjet, present within the large-R (top candidate) jet, is given by
\bea \label{Eqn:EnrgyFrac}
%\Delta E_{subjet}~=~\frac{E_{subjet}}{E_{topjet}}\\
{\Delta E_{T}^{subjet}~=~\frac{E_{subjet}}{E_{topjet}}}
\eea
Where, $E_{subjet}$= transverse energy of the subjet, $E_{topjet}$= transverse energy of the top candidate jet. In Figure \ref{Fig:17}, we present the energy fraction for leading and subleading b-tagged jets along with the c-tagged jet for the signal and the SM backgrounds. The figure on the left shows that predominantly, the 60\% energy of the large-R jet is carried away by the leading b-tagged jet for the signal events, while the same distribution manifests itself as a flat nature for the irreducible background. The energy fraction possessed by the leading b-tagged jet for the other background processes is characterized by the irregularities. Figure \ref{Fig:17} (center) shows that the sub-leading b-tagged jet for the signal events carries $\sim$ 20\% of the energy of the large-R jet, while the same fraction reduces to $\sim$10\% for the irreducible background. This implies somewhat asymmetric energy sharing among the (b-tagged) subjets. For the QCD background, the distribution peaks at $\sim$5\% and decays exponentially. The fraction of the energy carried by the c-tagged jet, shown in Figure \ref{Fig:17} (right), for the signal events peaks at $\sim$10\% and follows a slow exponential decay with a hump at $\sim$40\%. We have verified that the asymmetric energy distribution between the Higgs boson and the c-quark 
from the decay of top leads to such an observation. In contrast, all background processes, except the irreducible background, have a skewed distribution of the jet energy fraction of the c-tagged jet that peaks around 80\% of the topjet energy.

\begin{figure}[!htb]
\centering   % \begin{center}/ \end{center} takes some additional vertical space
\mbox{
{\includegraphics[width=.3\textwidth]{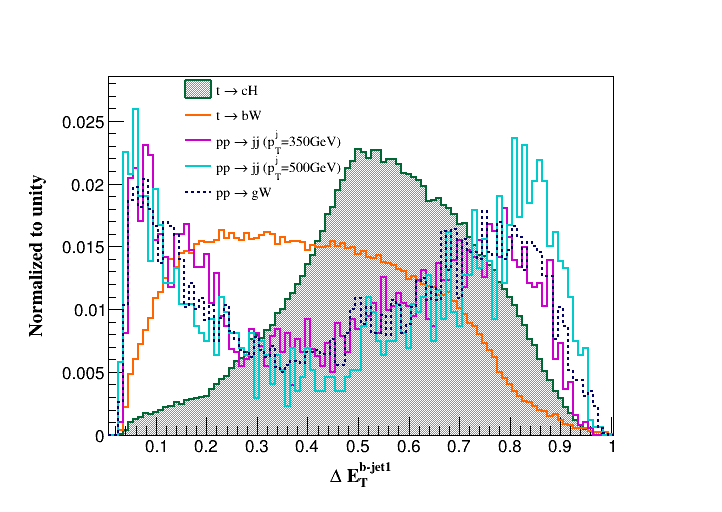}  }
{\includegraphics[width=.3\textwidth]{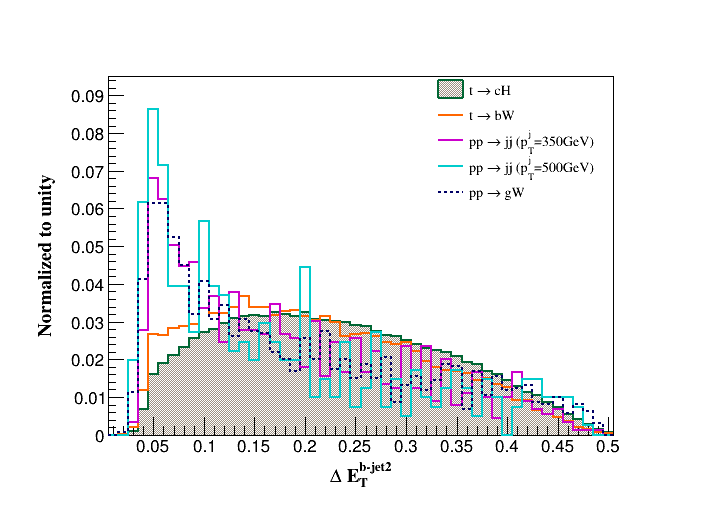}  }
{\includegraphics[width=.3\textwidth]{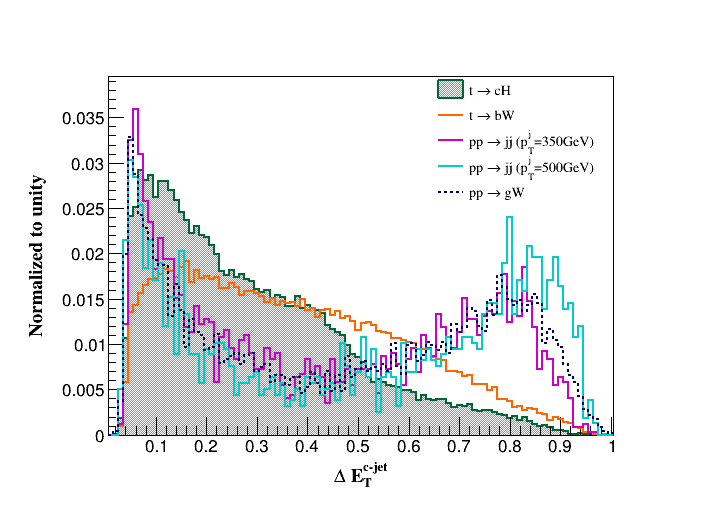}  }}
\caption{\label{Fig:17}Normalized distribution for the energy fraction of the leading b-tagged jet (left), subleading b-tagged jet (centre), and c-tagged jet (right) for signal and different SM background events.}
\end{figure}

%------------------------------------------------------------------------------------------
\item{\bf Fractional distribution of the invariant mass for different combinations of subjets:}

We further introduce another insightful observable, the fraction of the large-R jet mass acquired by different combinations of the subjets present within the boosted jet. The observables are defined as,
\bea  \label{Eqn:MassFrac}
\Delta X_{ij}  &=& \left( 1 - \frac{M_{ij}}{M_{topjet}} \right)  \nn \\
\Delta X_{ijk} &=& \left( 1 - \frac{M_{ijk}}{M_{topjet}} \right)
\eea 
where, $M$ refers to the (invariant) mass of the object, $M_{topjet}$ denotes the mass of the top-candidate jet, and $i, ~j, ~k$ are the indices to label the subjet structure within the large-R jet.  Equation \ref{Eqn:MassFrac} calculates the fractional mass of the large-R jet preserved after subtracting the invariant mass of two (or three) subjets inside it. The identification of a correct light jet is very important, and to obtain it we perform a $\chi ^{2}$-analysis \cite{ATLAS:2022ygn}. We retain only those light jets for which $\chi ^{2}$ is minimum, with $\chi ^{2}$ given by,
\beq
\chi ^{2} ~=~ \left( \frac{M_{bj} - M^{ref}_{h}}{\sigma _{M_{h}}} \right)^{2} ~+~ \left( \frac{M_{bjc}-M^{ref}_{top}}{\sigma_{M_{top}}} \right)^{2} ~+~ \left( \frac{(M_{bjc}- M_{bj}) - (M^{ref}_{top} - M^{ref}_{h})}{\sigma_{M_{top} - M_{h}}} \right)^{2}
\eeq 
where
$M_{bj}$: Invariant mass of b-tagged jet and light jet, $M^{ref}_{h}$: Reference mass of the Higgs boson = 125 GeV, $M_{bjc}$: Invariant mass of b-tagged jet, c-tagged jet, and light jet, $M^{ref}_{top}$: The reference mass of the top quark = 172 GeV, and all $\sigma$ s are set to 10 GeV. 

\begin{figure}[!htb]
\centering   % \begin{center}/ \end{center} takes some additional vertical space
\mbox{
{\includegraphics[width=.4\textwidth]{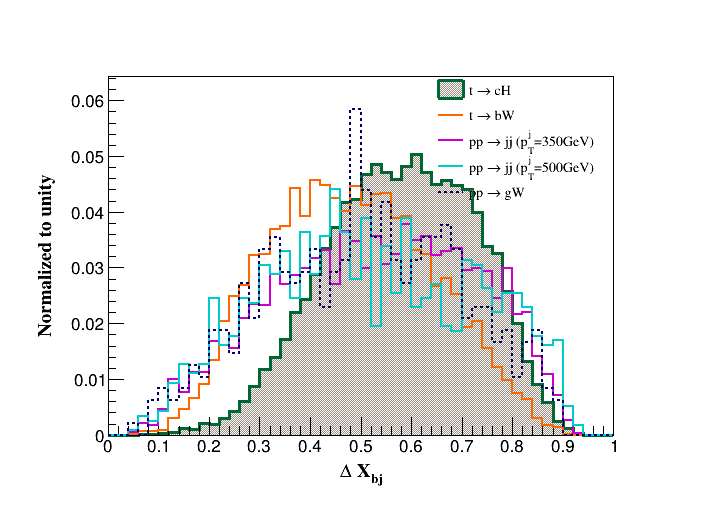}  }
{\includegraphics[width=.4\textwidth]{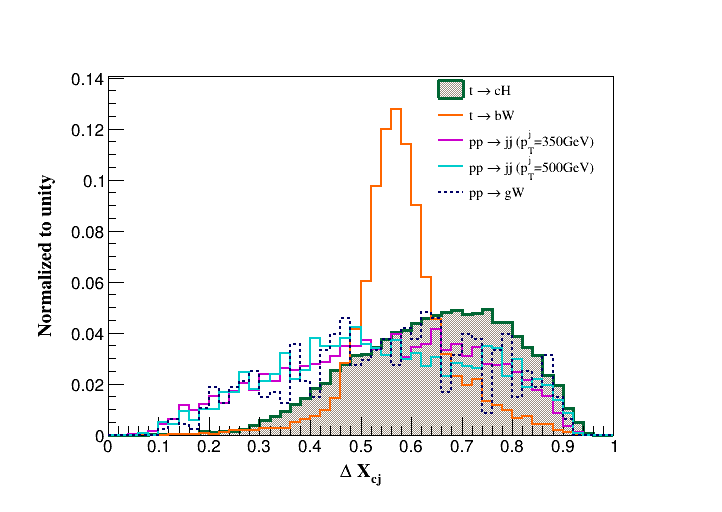}  } 
}
\caption{\label{Fig:18}\textit{Left}: Mass fraction of large-R, top-candidate jet retained after the subtraction of the invariant mass of b and the light jets, \textit{Right}: Mass fraction of large-R jet forsaken after considering the invariant mass of c and the light jets.} %The figure shows the mass fraction distribution for the Signal, and different SM Background processes.}
\end{figure}

Figure \ref{Fig:18} (left) indicates that the energy carried away by the b- and light jets for the signal is much smaller than that for the irreducible background. The reason behind such behavior stems from the fact that the light jet for the signal events appears principally from the radiative effects (or mistagging of heavy-flavor jets) whereas the background has an s-jet with sufficiently large $p_T$ to be labeled as light jet. The distributions for QCD events look as expected. The $\Delta X_{cj}$ distribution shows a sharp peak at $\sim$0.55 for the irreducible SM background; see Figure \ref{Fig:18} (right). With the s-jet identified as the light jet, the invariant mass of c and the light jets should peak around the mass of the $W$ boson ($\sim$ 80 GeV) and, therefore, $\Delta X_{cj}$ for such instances should be $\sim \left( 1 - \frac{80}{172} \right)$ which would lie within $\left[ 0.5, ~ 0.6 \right]$. The same distribution for the signal is highly skewed to the left with a peak around 0.75, indicating the invariant mass fraction of the large-R jet retained by the c and the light jet is $\sim$0.25, anticipated from the high asymmetry in top decay for the signal as discussed in the previous sections. The QCD background events display statistical randomness in the behavior. 

%Furthermore, we also calculate the fraction of the top candidate jet mass left after subtracting the invariant mass of the b, c and light jets, denoted as $\Delta X_{bcj}$ and in Figure \ref{Fig:18} (right). For the irreducible background it is found to be low, as expected. For the signal events, it peaks at a slightly higher value compared to the former as for signal light jet is appearing from the QCD radiations. The QCD backgrounds, with high cross-section and negligible possibilities of having b, c and a light jets simultaneously do not show any insightful pattern.

%----------------------------------------------------------------------------------------
\end{itemize}

Before we discuss the results, it is important to introduce the trigger needed to select these events. The presence of a boosted top jet implies that the events we consider will consist of high missing transverse energy ($\cancel{E}_{T}$ > 200 GeV) and a charged lepton with $p_{T} > 200$~GeV generated from the decay of the W boson. The charge of the lepton indicates the production of top or anti-top at the particle level, and the high $\cancel{E}_{T}$ can be set as a trigger to minimize spurious SM backgrounds. 
%%%%%%%%%%%%%%%%%%%%%%%%%%%%%%%%%%%%%%%%%%%%%%%%%%%%%%%%%%%%%%%%%%%%%%%%%%%%%%%%%%%%%%%%%%%%%

%---------------------------------------------------------------------------
%\subsection{Comparison with Conventional cut-based Top Taggers}
\subsection{Performance of the Conventional cut-based Top Taggers}

Various top-tagging algorithms \cite{Plehn:2011tg} have been proposed so far to efficiently tag large-R jets reconstructed from a boosted top using the kinematics of the jet and its substructures. The two predominantly used toptaggers, namely John Hopkins Top Tagger (JHTopTagger) \cite{JHToptagger} and HEPTopTagger \cite{Plehn:HepToptagger} are considered here for comparison. Both taggers identify a boosted top quark from its decay to $b W$ followed by the hadronic decay of the $W$ boson. Implementation of JHTopTagger for our event topologies enables the identification of top jets with an efficiency of 32\% for the signal and 43\% for the irreducible background $t \to bW$. The tagger identifies a top jet in the 7\% cases for the collective QCD background. Upon replacing the JHTopTagger with the HEPTopTagger, the efficiency further drops to 18\% for the signal and 35\% for the irreducible background, while the same for the collective QCD background reduces to 3\%.

As the observation suggests, conventional toptaggers do not perform well for the identification of the top jets of signal events from their exotic decays to the $c, ~b$ and $\bar{b}$ modes. The reason behind their poor performances is rooted in the fact that these taggers are designed and fine tuned for the identification of large-R jets from a boosted top decaying to $b W$ where the $W$ further undergoes hadronic decay to an up- and a down-type (anti)quarks, whereas our signal events principally comprises of top decaying to $c H$ with $H$ leading to $b \bar{b}$ pair. One possible resolution could be to modify different parameters of the HEPTopTagger to formulate the algorithm around the Higgs boson mass, instead of the $W$ mass. Following the suggestions of Ref. \cite{Greljo:2014dka}, we alter some of the parameters necessary to capture the important regions in the mass planes constructed using different invariant mass variables, and found that the contribution of the irreducible background ($t \to bW$) reduces significantly due to the demand of two b-tagged jet inside, however negligible impact was observed for signal events. Therefore, it is important to go beyond the cut-based taggers which are primarily devised for hadronically decaying top quark, and build a tagger based on more relevant jet-substructure based observables followed by a multivariate analysis.

%%%%%%%%%%%%%%%%%%%%%%%%%%%%%%%%%%%%%%%%%%%%%%%%%%%%%%%%%%%%%%%%%%%%%%%%%%%%%%%%%%%%%%%%%%%%%
%%%%%%%%%%%%%%%%%%%%%%%%%%%%%%%%%%%%%%%%%%%%%%%%%%%%%%%%%%%%%%%%%%%%%%%%%%%%%%%%%%%%%%%%%%%%%
%\subsection{MultiVariate Analysis} \label{SubSec:TMVA}
\subsection{Performance of Machine Learning based Top Taggers}\label{SubSec:TMVA}

We have, by far, discussed the observations we made till the detector-level simulation of boosted, collimated candidate topjets with large radius. Our objective is to narrow down and propose a tagger to identify a rare top decay as a probe of New Physics. Basically, the crux and essence of our study is, at its heart, a classification problem: to come up with an algorithm that proves to be effective over the existing ones in identifying and segregating events involving decay of a boosted top to Higgs boson and a charm quark, encapsulated in a large-R jet. In order to accomplish our goal, we construct different physical observables, or features, to segment the feature space, built with each feature at a time, into blocks and to identify the blocks with enhanced presence of signals over the SM backgrounds. In contrast to a standard cut-flow method, the MultiVariate Analysis (MVA) methods consider all the features in a single go and obtain regions with enhanced signal density over the background. Therefore, the new features generated by the MVA algorithms often possess an enhanced classification ability over traditional cut-based methods. We use the {\tt scikit-learn} library \cite{scikit-learn} to implement different MVA algorithms. Events that satisfy the selection criteria for large-R jets are selected for the MVA.  

%ROOT Library, "Toolkit for Multivariate Analysis" (TMVA) \cite{hoecker2007tmva} 

%mass and $p_T$ of the candidate large-R jet containing atleast one b- and c-tagged jets are only selected for the MVA. 

%%%%%%%%%%%%%%%%%%%%%%%%%%%%%%%%%%%%%%%%%

The event weights play an important role in Multivariate analysis. In Table \ref{Tab:2}, we show the expected number of signal and background events, denoted by $N_{Exp}$, for an integrated luminosity of 3 ab$^{-1}$ at the High-Luminosity LHC (HL-LHC) \cite{HLLHC2}. The quantity $N_{Exp}$ is defined as $N_{Exp}$ = $\frac{N_{precut}}{N_{MC}} \times \sigma_{MC} \times \mathcal{BR} \times (\int \mathcal{L} dt)$, where $N_{Precut}$ denotes the number of events satisfying basic selection cuts. The notation used to label different processes follows Table \ref{tab:simulation}. The quantity $\sigma_{MC}$ denotes the production cross section for the processes, while $\mathcal{BR}$ is the net branching fraction of the combined decay channel where we use $\mathcal{BR} \left( W^- \to \mu^- \bar{\nu}_{\mu} \right) = 0.11$, $\mathcal{BR} \left( H \to b \bar{b} \right) = 0.6$, and $\mathcal{BR} \left( W \to c \bar{s} \right) = 0.32$ \cite{PDG}. For each process, except for Background 2, we simulate $N_{MC} = \mathcal{O}(10^6)$ events following the procedure discussed in Section \ref{Sec:App1}. Due to the high production cross section, we generate $\mathcal{O}(10^7)$ events for Background 2. For signal events, we set the branching fraction for $ t \to c H$ to be 100\%, which we later vary to compare with the current experimental limits (see Table \ref{Tab:signi}). Table \ref{Tab:2} includes an important quantity, called Weight, associated with individual processes. It is defined as weight = $\frac{1}{N_{MC}} \times \sigma_{MC} \times \mathcal{BR} \times (\int \mathcal{L} dt)$. The performance of the MVA methods, in principle, depends on the weights as it is proportional to the production cross section. Although, in principle, we can use all the background processes; however, we find that QCD events play a merginal role in the classification performance. In fact, we can always minimize the effect of the QCD background events by applying the trigger discussed above. 

%---------------------------------------
\begin{table}[!htb]
\centering  
\hspace*{-0.25cm}
\begin{tabular}{|c|c|c|c|c|c|}    \hline
\textbf{Process} & \textbf{$\sigma_{MC}$} & \textbf{$\mathcal{BR}$} & \textbf{$N_{Precut}$} & \textbf{$N_{Exp}$} & \textbf{Weight} \\
 & (in fb) &  &   & (for $(\int \mathcal{L} dt) = 3 ab^{-1}$) &  \\   \hline  \hline
Signal & 435 & ${6.6 \times 10^{-2}}$ & 79036 & $6.8 \times 10^3$ & $8.6 \times 10^{-2}$  \\ \hline
Background 1 & 435 & ${3.5 \times 10^{-2}}$ & 88724 & $4.1 \times 10^{3}$ & $4.6 \times 10^{-2}$  \\   \hline
%%%
Background 2 & $3.91 \times 10^6$ & {1} & 6302 & $1.3 \times 10^7$ & $2.1 \times 10^3$  \\   \hline
%%%
Background 3 & $6.08 \times 10^5$ & {1} & 2394 & $4.3 \times 10^6$ & $1.8 \times 10^3$  \\   \hline
%%%
Background 4 & $2.04 \times 10^3$ & 0.11 & 841 & 566 & $6.7 \times 10^{-1}$  \\   \hline
\end{tabular}
\caption{\label{Tab:2}Table portraying the expected number of statistics for the signal and the SM Backgrounds generated at Leading Order for an integrated luminosity of 3 ab$^{-1}$. Note, the quantity ${\mathcal {BR}}$ is the net branching fraction of the combined decay channel for each processes. }
\end{table}
%---------------------------------------
%%%%%%%%%%%%%%%%%%%%%%%%%%%%%%%%%%%%%%%%%%
Having discussed the weights for the different processes, we move on to identify features (variables or physical observables relevant for the classification) pertinent to the segregation of the events into signals and backgrounds. Now, most state-of-the-art toptaggers do not use flavor tagging at the subjet level like b- and c-tagging when analyzing large-R jets. Therefore, for a fair assessment and comparison with other methods based on detector level observables, we introduce two sets of input variables, namely Set 1 and Set 2. In Set 1, we consider the three leading subjets of the large-R jet and construct the input features (observables) using the information shared among these subjets (untagged), while in Set-2 we use the flavor-tagged subjets to calculate the same set of quantities (see Table \ref{Tab:3}). Furthermore, we also include a few observables based on the displaced tracks and vertices in Set 2; for details on the construction of displaced tracks and vertices, we refer to Appendix \ref{Sec:App3}. We consider the subset of the data samples for signal events ($\rm t \to cH$, $\rm H \to b\bar{b}$) and irreducible SM background events ($\rm t \to bW$, $\rm W \to c \bar{s} $) that include at least one b- and c-tagged subjets within the top-candidate large-R jets.%\footnote{The QCD events can be tackled using the trigger discussed in the preceding sections.} 
We perform a 80\%-20\% split of the dataset containing event-by-event information for the signal and the irreducible SM background, and train a XGBoost Classifier. The performance of the classifier is then tested in the remaining 20\% data set. The accuracy score, which represents the percentage of correctly classified events, is 85\%. 

\begin{figure}[!htb]
\centering  
{\includegraphics[width=.45\textwidth]{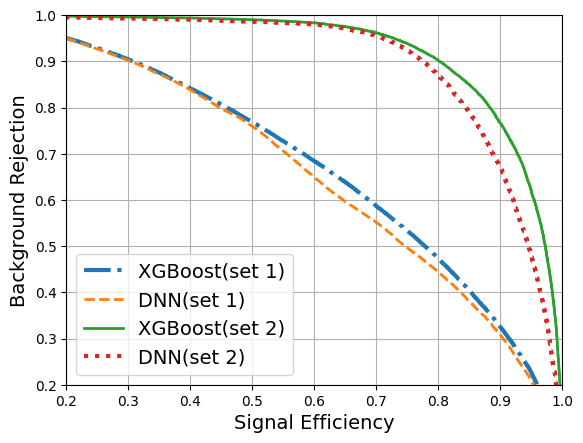}  }
\caption{\label{Fig:21}Comparison of the performances of the classifiers built using  XGBoost and Deep Neural Network (DNN) based tagger.}
\end{figure}

The performance of the Extreme Gradient Boosting (XGBoost) algorithm \cite{chen2016xgboost} for the two input sets is shown in Figure \ref{Fig:21} using the ROC curve obtained after optimizing the identification of signal events over the misidentification rates of background events. %\footnote{More details on the family of boosting techniques and their comparison, we refer \cite{chen2016xgboost, weak-to-strong, friedman2001greedyGB, Zhang2010}} 
 The dashed-dotted blue and solid green lines represent the two cases for Set 1 and Set 2 respectively. The improvement in the background rejection rates, and thereby enhancing signal efficiencies, is observed while using high-level features based on b- and c-tagged jets. In fact, we observe that the fraction of the invariant mass formed with c-tagged and light subjets, and with a b- and a c-tagged subjet within the large-radius jet along with the energy sharing among the b- and c-tagged subjets are the prominent contributors to the classifier. The role of the subjettiness variables, different features on the event kinematics, and observables based on displaced tracks information comes later in the classification purpose. For comparison, we also analyze the performance of the MVA method when equal weights are provided for all the processes. The results seem to be marginally affected. We also progressed one step further to compare the performance of the model we built using XGBoost with a model constructed using another boosting algorithm, Adaboost \cite{schapire2013explaining}. The XGBoost classifier has a higher background rejection rate compared to AdaBoost due to its better regularization, parallel processing capabilities, and gradient optimization\footnote{The classifying capability of a trained model is often expressed in terms of a Confusion matrix written in terms of true positive, true negative, false positive and false negative predictions. For the trained model that uses the features of Set 2 and the irreducible background only, we find that the XGBoost classifier has an accuracy of 85\%, with a precision of 87\% and a recall of 80\%. These quantities indicate that the classifier built here is both efficient and robust.}.

% ------------------------------------
\begin{table}[!htb]
\centering
\begin{tabular}{c}   \hline
\textbf{Input variables for MVA} \\   \hline  
\begin{tabular}{c|c}
%\multicolumn{}{}{}
  {\bf Set 1}    & $N_{Subjet}$, $p_{T}^{j_1}$, $p_{T}^{j_2}$, $\tau_{21}^{1}$, $\tau_{32}^{1}$, $\Delta X_{12}$, $\Delta X_{13}$,  $\Delta X_{23}$, $\Delta E_T^{j_1}$, $\Delta E_T^{j_2}$ \\ [2mm]
\hline
 {\bf Set 2}    & $N_{Subjet}$, $p_{T}^{b-jet}$, $p_{T}^{c-jet}$, $\tau_{21}^{1}$, $\tau_{32}^{1}$, $\Delta X_{cj}$, $\Delta X_{bc}$, $\Delta X_{bj}$, $\Delta E_T^{b-jet1}$, \\
 & $\Delta E_T^{b-jet2}$, $\Delta E_T^{c-jet}$, $N_{light-jet}$, $N^{Displaced}_{Track}$, $M_{DV}$, $N_{DV}$ \\ [2mm]
\hline 
\end{tabular}
\end{tabular}
\caption{\label{Tab:3}{The two sets of variables used as input for MultiVariate Analysis. See the text for details. 
%The Set 1 is the variable set for the untagged case and set 2 is the variable set while b and c tagging is considered .
}}
\end{table}
%
%%%%%%%%%%%%%%%%%%%%%%%%%%%%%%%%%%%%%%%%%%
%%%%%%%%%%%%%%%%%%%%%%%%%%%%%%%%%%%%%% 

In addition, we have also compared the performance of our tagger with that of a model built with Deep Neural Network (DNN) \cite{Cun:DNN,  schmidhuber2015deep}. The irreducible background of SM, among the four possible ones, is considered for this comparative study. We provide the same dataset, used to train the XGBoost and Adaboost models, as input, into an architecture comprising of two 1-dimensional convolutional layers containing 16 filters, with a kernel size of 4. Four dense layers, each containing 100 nodes, follow the convolution layers and precede the output layer, containing one perceptron, densely connected with the previous hidden layer of the network. The Leaky ReLU activation function \cite{LReLU} in all hidden layers keeps the architecture free from the vanishing gradient problem during back-propagation. The output layer uses the sigmoid activation function to classify inputs as signal or background. The entire dataset, with around 500k events comprising both the signal and background, has been divided into a 4:1 ratio to train and test the DNN model. The DNN architecture has been created using Keras \cite{chollet2015keras}, and the performance of the model has been visualized with an ROC curve using the scikit-learn \cite{scikit-learn} library. The tagger we propose with XGBoost, outperforms the DNN classifier, especially in the higher signal efficiency regions for Set-2 (see the orange dashed and red dotted lines in Figure \ref{Fig:21}). 

%The former optimises to a signal efficiency of 80\% against the background mis-identification rate of 12\%, while for the latter the background mistagging rate is 16\% for the same signal efficiency}

In addition to those high-level observables constructed using jet substructure information, we also study the prospect of the \textit{Jet Image} technique constructed using the $p_T$, $\eta$, $\phi$ information of the jet constituents. We prepare the image of the top-candidate jet for each event, considering 3,00,000 events for both signal and background, and break the entire dataset into 7:3 ratio for training and testing purposes. We fit a CNN network similar to Ref.\cite{Kasieczka:2017nvn} having two convolutional blocks each containing two convolutional layers (with 8 filters for each layer in the first block and 4 filters for the same in the second block), with kernel size $2 \times 2$ separated by a Max Pooling layer of dimension $2 \times 2$. The output was then passed through three dense layers of 128 nodes. The ReLU activation function is used in all the layers mentioned above. Finally, a dense layer consisting of 2 nodes with Sigmoid activation function gives the final output. As the jet images of the Signal and the irreducible background events with a three-prong top decay structure look similar, the classifier achieves around 60\% signal efficiency over 30\% background efficiency. 

%%%%%%%%%%%%%%%%%%%%%%%%%%%%%%%%%%%%%%
%\iffalse
Finally, motivated by the findings of \cite{Kasieczka:2019dbj} and the comparisons presented in Tables 1, 2 in \cite{Gong:2022lye}, we compare the performance of the XGBoost tagger with the Lorentz equivariant graph neural network, namely {\tt LorentzNet}, a state-of-the-art framework for jet tagging based on the deep learning model that preserves Lorentz symmetry \cite{Gong:2022lye}. The design of LorentzNet is based on the universal approximation theorem, which guarantees both the equivariance and universality of the network. Following the proposal of \cite{Gong:2022lye}, the input features include the 4-momentum of the jet constituents along with the $p^{jet}_T$, mass of the jet constituents, and the 4-momentum of the top quark at the truth level. The jet with the maximum number of constituents ($N_{\rm cons}^{\rm jet}$) is identified by analyzing all the events, and each of the jets having the number of constituents less than $N_{\rm cons}^{\rm jet}$ is zero-padded. Note that the LorentzNet architecture uses jets as graphs defining constituent particles as nodes, where the 4-momentum of the particles are used as coordinates in the Minkowski space.
%The feature information, the mass of the constituents, is used as a scaler which is different for the different nodes in the graph whereas the $p^{jet}_T$, truth level 4 momentum of the top holds same values for all the nodes (constituents) in the jet.

We simulate around 1.2 million samples and divide the data set into 60:20:20 ratios for training, validation, and testing purposes, respectively. After successfully training the model, it achieved 63\% accuracy with an AUC of around 65\% for 35 epochs. In addition to the input features of the original setup, we next extend the input parameters with the high-level objects mentioned in Table \ref{Tab:3}. However, the AUC/accuracy did not show a significant improvement.\footnote{We validated our LorentzNet implementation by performing the top vs QCD classification where the top decays through the dominant mode of the SM and $p_T$ of the candidate jet lies within [400 - 500] GeV window. We obtain 95\% AUC with an accuracy of 98\% for 20 epochs, which is in good agreement with the original paper.} Our observation complies with that made in \cite{Sahu:2024sts}, and indicates that the 4-vectors of the respective jet constituents of signal and background events, which in this context are closely resembling ($t \to cH, ~H \to b\bar{b}$ and $t \to bW, W \to c\bar{s}$), largely controls the classification ability of LorentzNet, marginalising the contributions of the high level parameters, which in this case, play the pivotal role. Table \ref{tab:toptagger}, provides a comparison between different top-tagging architectures (first column), and respective AUCs which is a measure of the ability of the classifier to distinguish the signal over the backgrounds (second column), and the background rejection rate at the signal efficiency 60\% (${\rm Rej}_{60\%}$). The performance of CNN and LorentzNet frameworks, compared to the taggers based on high-level features viz. DNN and XGBoost, can be reconciled by looking into the internal structure (both energy profile and flavor) of the boosted large-R jet under consideration. The information stored at the constituent level for the top decay to $cH$ mode overlaps significantly with that for the top decay to $bW$. Hence, the limitation of the top taggers based on low-level features (similar to Set 1, without flavor tagging) in efficient identification of the FCNC decays of top necessitates devising a dedicated top tagger which includes high-level features using the jet substructure information and flavor tagging among the subjets (viz. observables of Set-2). However, it would be interesting to study the sensitivity of a two-stream convolutional neural network architecture (and its variants) \cite{Lin:2018cin} that combines the information retrieved from the boosted jet constituents and high-level features, focusing principally on three-prong jet substructures. A recent advancement in multi-level feature-aggregated transformer \cite{10095095} constructed on the edifice of global content and structure aggregation and customized local convolution aggregation to analyze variables based on the high-level jet substructure and the image capturing the dynamics of jet constituents can be a future directive for a potentially powerful classifier to detect FCNC decays of a boosted top.

%\iffalse
\begin{table}[!tbh]
    \centering
    \begin{tabular}{|c|c|c|}
    \hline
 %    &  \\ 
      Architecture & AUC  &  ${\rm Rej}_{60\%}$ \\ [2mm]
      \hline 
      AdaBoost & 0.65 (0.911)   & 2.56 (53)\\ %.93
      XGBoost & 0.71 (0.93) & 3.225(65) \\  %.95
      DNN &   0.69 (0.913)& 2.94(54)\\   %0.91
      CNN &0.73 & 3.33 \\     %0.7
      LorentzNet& 0.66 & 2.56 \\  %.61
    \hline
    \end{tabular}
       \caption{Performance comparison between different tagging algorithms, showing background rejection rate ($\frac{1}{\epsilon_B}$) at 60\% signal efficiency with $\epsilon_B$ being the background efficiency. The numbers in the parenthesis represent the case when the input features of Set-2 are considered.}
    \label{tab:toptagger}
\end{table}
%\fi

%%%%%%%%%%%%%%%%%%%%%%%%%%%%%%%%%%%%%%%%%%%%%%%%
As already mentioned, the FCNC interactions of the top quark and the Higgs boson through the $t-c-H$ coupling play an important role in our quest for physics beyond the SM. Several searches for this anomalous coupling have been carried out by the ATLAS and CMS collaborations at the LHC. Analyzing final states containing exactly one lepton (muon
or electron) and at least three jets, among which at least two are identified as b-jets, the upper limit in $BR( t \to cH)$ is 0.094\% at the 95\% C.L. \cite{CMS:2021gfa}. Considering other possible final states but with the Higgs boson decaying to $\tau$ -leptons or photons, the upper limit is $BR( t \to cH) < 0.094\%$ \cite{ATLAS:2022gzn} and $BR( t \to cH) < 0.073\%$ \cite{CMS:2021hug} at the 95\% C.L., respectively. To assess the prospect of probing this exotic mode in the boosted regime through the proposed tagging algorithm, we calculate the signal significance, defined as 
$\mathcal{S}  = \sqrt{2 \left[ (S + B) \ln\left(1 + \frac{S}{B}\right) - S \right]}$  \cite{Cowan:2010js} with S and B being the number of signal and background events that satisfy the jet selection cuts and also pass the trigger requirement.%that has a high $p_T$ lepton and missing transverse energy $\cancel{E_T} > 200$~GeV. 
As evident in Table \ref{Tab:signi}, the proposed toptagger can exclude $BR( t \to cH) \sim 0.5\%$ with a 95\% C.L. with 150 ${\rm fb}^{-1}$ luminosity at the 13 TeV run of LHC. To further quantify the impact of the proposed tagger on new physics, we use the Asimov estimate of statistical significance \cite{Cowan:2010js}. An important feature of this kind of significance measure is that it allows for the inclusion of systematic uncertainties. The reach of new physics changes marginally with the inclusion of the systematic uncertainty of 5\% in the background estimation. This exercise suggests that with 3 ab$^{-1}$ of collision data, our method would provide a signal significance of 1.5$\sigma$ for $BR( t \to cH) = 0.1\%$, too minimal to claim a discovery, but worthy of redirecting future searches considering other production modes of the top quarks and both b- and c-tagging at the subjet level, along with sophisticated search strategies at the higher luminosity runs of the particle collider.

%========================
\begin{comment}
\begin{table}[!htb]
\centering
\begin{tabular}{|c|c|c|c|c|}
\hline
{Luminosity ($fb^{-1}$)} & Flavor tagging & $BR( t \to cH)$  & \(\mathcal{S} \,\) & \(\mathcal{S} \, (\sigma_b = 0.05 \cdot B)\) \ \\\hline
150 &1b1c &0.5\%  &  1.6 & 1.5 \\
& &0.1\%  & 0.33 & 0.32 \\
\cline{2-5}
& 2b&0.5\% & 0.74 & 0.64  \\ 
& &0.1\% & 0.15 &  0.13 \\ \hline
3000 & 1b1c & 0.5\% & 7.02 & 5.43 \\
& &0.1\% & 1.47 & 1.18 \\ 
\cline{2-5}
& 2b &0.5\% & 3.29 & 1.86  \\
& &0.1\% & 0.66 & 0.38  \\ \hline
\end{tabular}
\caption{Comparison of signal significances for two different choices of flavor tagging at the subjet level at 150 and 3000 $fb^{-1}$ integrated luminosities.
}
\label{Tab:signi}
\end{table}
\end{comment}

%=========================
%========================
\begin{table}[!htb]
\centering
\begin{tabular}{|c|c|c|c|}
\hline
Luminosity  & Flavor tagging & $BR( t \to cH)$  & \(\mathcal{S} \,\) \\ 
(${\rm fb^{-1}}$) & (of subjets) &  & \\\hline
150 &1b1c &0.5\%  &  {\bf 1.6}  \\
& &0.1\%  & 0.3  \\
\cline{2-4}
& 2b&0.5\% & 0.7   \\ 
& &0.1\% & 0.2  \\ \hline
3000 & 1b1c & 0.5\% & 7.0  \\
& &0.1\% & {\bf 1.5}  \\ 
\cline{2-4}
& 2b &0.5\% & 3.3   \\
& &0.1\% & 0.7   \\ \hline
\end{tabular}
\caption{Comparison of signal significances for two different choices of flavor tagging at the subjet level at 150 and 3000 ${\rm fb}^{-1}$ integrated luminosities.
}
\label{Tab:signi}
\end{table}

We would like to end this section by pointing out that the approach followed in this work is generic, optimizing the tagging of a boosted top jet with a large radius, ideally searching for a three-prong structure inside the large-R jet, and efficiently tagging at least one b- and one c-tagged jet among the observed subjet structures. For cases with two b-tagged jets, it also affirms the origin of these jets from Higgs boson decay after the reconstruction of the di-jet invariant mass. However, note that the prescription followed here is much robust and can be implemented with minor modifications for similar probes of collimated large-R jets involving any flavor changing top decays \cite{Gutierrez:2020eby}, for the probes of R-parity violating Supersymmetric theories \cite{Barbier:2004ez}, and in identification of the vector-like quarks \cite{Cetinkaya:2020yjfvlq}. 
\section{Interpretation of Model Features using SHAP Analysis}  \label{Sec:Shap}

The ROC curve, as discussed in Section \ref{SubSec:TMVA}, through a trade-off between signal identification and misidentifications, provides a measure of model performance, as the threshold for classification varies. This section provides a measure for the contribution of the observables to the classification, using Shap (SHapley Additive exPlanations) \cite{ShapBook}, a notion borrowed from the cooperative game theory \cite{2022arXiv220205594R}. The Shapley value, determined for the $i^{th}$ feature, given by,
\begin{equation} \label{Eqn:ShapVal}
\displaystyle \varphi _{i}(v) ~=~ \sum _{S\subseteq N\setminus \{i\}} ~ {\frac {|S|!\;(n-|S|-1)!}{n!}} ~ \left( v(S\cup \{i\}) ~-~ v(S) \right),
\end{equation}
where,
\vspace*{-0.2cm}
\bea
S &:& \text{subset of the total number of features $n$, removing the $i^{th}$ one}  \\  \nn
\lvert S \rvert & : &\text{cardinality of the set S}   \\  \nn
v & : & \text{a function which maps the subset of features to an integer number}. 
%(in this case, a performance metric of the classifier).}
\eea 
It quantifies the contribution of the $i^{th}$ feature, from the weighted sum of the difference in classifier performance, when that feature is incorporated versus when it is not, considering all possible sets of features (S) eliminating the $i^{th}$. This, in turn, enables the selection of pronounced features for the classification and thus improves the classifier efficiency. 
 
%%%%%%%%%%%%%%%%%%%%%%%%%%%%%%%%%%%%%%%%%%%%%%%%%%%%%%%%%%%%%%%%%%%%%%%%%
We estimate the contribution of each feature to signal background segregation, using the Tree Explainer \cite{land.lee}. Figure \ref{Fig:1sp}, a layered violin plot \cite{violplt}, represents the summary of the XGBoost classifier. Each point on the summary plot represents a Shapley value for a given feature in an event, subsequently smoothed to display the global distribution of the Shapley value corresponding to that feature. The y-axis of this plot represents the features in relevance order. The x-axis, for a certain feature, represents the Shapley value for the feature concerned for the given collision event. The color code indicates the relative magnitude of the feature, ranging from low to high values, represented, respectively, by blue and red. We make the following observations from the Figure \ref{Fig:1sp}: 

%----------------------------------------------------
\begin{figure}[!htb]
\centering 
\includegraphics[width=0.60\textwidth]{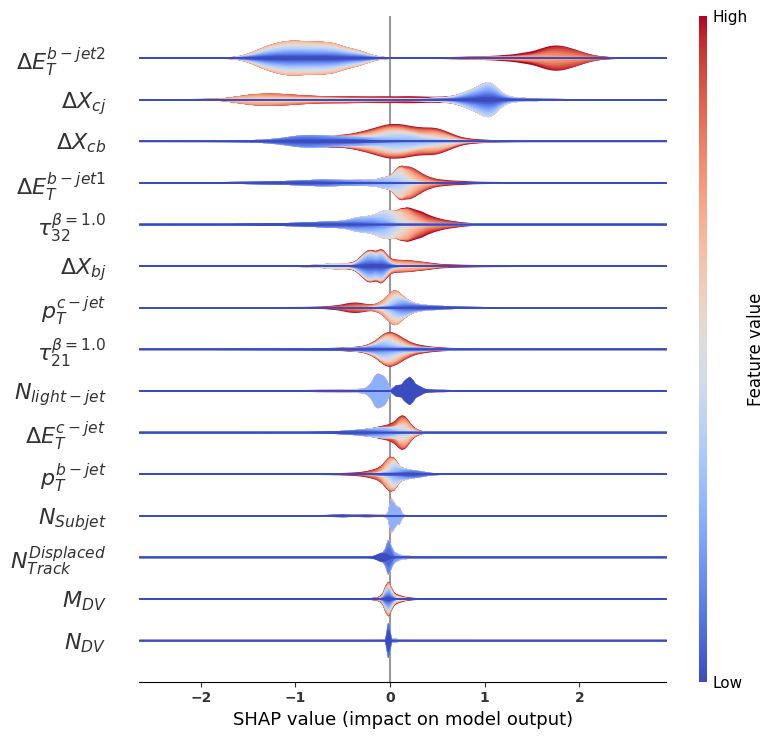}%summary_7aug.png
\caption{\label{Fig:1sp} Summary plot with Tree Explainer, portraying contribution of each feature, for the XGBoost Classifier, to the appropriate predictive capacity of the model.}
\end{figure}
%---------------------------------------------------- 
%%%%%%%%%%%%%%%%%%%%%%%%%%%%%%%%%%%%%%%%%%%%%%%%%%%%%%%%%%%%%%%%%%%%%%
%\fi
\begin{itemize}
\item ${\rm \Delta E_T^{b-jet2}}$: The most important variable to classify the events whether to be signal-like or background-like is the fractional transverse energy of the sub-leading b-tagged jet. As expected,  the presence of an additional b-tagged jet inside the large-R jet is a key difference between the signal and the irreducible background. 
\item $\rm \Delta X_{cj}$: The second most decisive feature for the classifier, is the residual fraction of the invariant mass, $\rm \Delta X_{cj}$. For the background, the light jet is predominantly an s-jet originating from the $W$-decay and hence their invariant mass peaks around the resonant mass of W boson. Whereas for the signal it mostly appears from the QCD radiation effects of the heavier quarks or from the mistagging of the b or c-subjets for which $\rm \Delta X_{cj}$ is relatively higher. Therefore, as depicted in the plot, events with lower (feature) values of $\rm \Delta X_{cj}$ are predominantly classified as the signal, while events with higher values of $\rm \Delta X_{cj}$, in general, are classified mostly as background.
\item $\rm \Delta X_{cb}$: The third salient feature for the signal-background segregation is residual invariant mass fraction of one c- and one b-tagged subjet pair ($\rm \Delta X_{cb}$). Due to the absence of high asymmetry of mass between the top-decay products for the background, the residual mass fraction is minimal, and therefore these events are (dominantly) projected in the region of negative shaply values. 
\item${\rm \Delta E_T^{b-jet1}}$: Next in the hierarchy of importance, comes the feature referring to the fractional energy of the leading b-tagged jet (see Equation \ref{Eqn:EnrgyFrac}), present inside the large-R  jet. For the signal events, the leading b-tagged jet with higher $p_{T}$ originates from the decay of Higgs boson produced in association with a c-quark from top, and carries most of the energy of the top quark due to high asymmetry in mass. On the other hand, for the background, b-tagged jet appears from the top-decay along with W. Due to the mass asymmetry between b-quark and W, in this case, W carries most of the top energy, resulting in the energy of the b-tagged jet to take smaller values. This behavior is well reflected in the segregation of signals from the irreducible background. % events with higher value of the fractional energy carried by the leading b-tagged jet are classified as signal, while the events with smaller values are predicted to be  background.}
\item $\rm \tau_{32}^{\beta=1}$: This feature gives an insight of whether the large-R jet under consideration has a substructure which resembles more 3-prong like than 2-prong like, for which the value of the expression would be smaller. Even though both the signal and the irreducible background respectively lead to large-R jets comprising principally of three subjets, the Shapley approach, however succeeds in retrieving subtler contrast, showing events with higher value of $\rm \tau_{32}^{\beta=1}$ to be signal, while those with lower values, classified as the background. The effect of enhanced QCD radiation for the signal, in comparison to that for the background, addresses this apparent (counter-intuitive) behavior.
\item The explainability for the behavior of the rest of the features, arranged in decreasing order of their respective importance, in segregation between the signal and the background events, are preserved. We refrain from the detailed explanation for the rest of these features, they follow trivially like the ones stated above. Additionally, studying the SHAP dependence plot is valuable for visualizing how a specific feature, along with its interactions with other features, impacts the trained model's predictions. For more details, readers are referred to the Appendix \ref{Sec:App-dependence}.

%In the following part of this section, we discuss the contribution of some of the important features incorporated in the classifier model, and analyse some of them individually in connection with other features accounted in the classifier, on the predictability of the events.

%%%% SHAP values show how each feature affects each final prediction, the significance of each feature compared to others, and the model's reliance on the interaction between features.

%%% (previous text) 
%The explainability for the behavior of the rest of the features, arranged in decreasing order of their respective importance, in segregation between the signal and the background events, are preserved. We refrain from the detailed explanation for the rest of these features, they follow trivially like the ones stated above. However, we spend more time in digging into the explanation for the behavior of the pronounced features, placed at the top of Figure \ref{Fig:1sp}, which play decisive roles in the classification. In the following part of this section, we discuss the contribution of some of the important features incorporated in the classifier model, and analyse some of them individually in connection with other features accounted in the classifier, on the predictability of the events.

\end{itemize}
%%%%%%%%%%%%%%%%%%%%%%%%%%%%%%%%%%%%%%%5
%\begin{figure}[!htb]
%\centering 
%\includegraphics[width=0.50\textwidth]{figures/best7_roc.png}
%\caption{\label{Fig:2sp} Receiver Operating Characteristics or ROC curve for the model, built with seven most important features for classification. It quantifies and displays the model performance in appropriately classifying the signal and the background events. Blue line represents the ROC behavior for the model built with the first seven features enlisted in Figure \ref{Fig:1sp}, which registers 80\% signal efficiency in exchange of the background rejection of 15\%. The orange line, taking into account all 13 features, shows 80\% signal efficiency for the background rejection of $\sim$12\%.}
%\end{figure}
%\sout{We start with, the above-explained seven most important feature to build the same classification model, and check its performance in proper classification using the ROC plot. We find that the first seven features from the top, enlisted in Figure \ref{Fig:1sp}, enables the classification, almost as good as in the case where all of thirteen features are taken into account. This indicates that the first seven features carry the lion-share of the classification power of the model. The remaining six of the thirteen features account for the enhanced background rejection rate by $\sim$3\%, for the same value of the signal efficiency.}
%%%%%%%%%%%%%%%%%%%%%%%%%%%%%%%%%%%%%%%%%%%%%%%%%

%%%%%%%%%%%%%%%%%%%%%%%%%%%%%%%%%%%%%%%%%%%%%%%%%%%%%%%%%%%%%%%%%%%%%%%

Having presented an overview of the analysis on the individual and joint contributions of the features towards the segregation of events into signal and background, we next take up the task of proper identification and misidentification of the events locally. 

%As already stated, SHAP values reveal the impact of each feature on the final prediction, highlight the relative importance of each feature compared to the others, and demonstrate the model's dependence on feature interactions. Therefore, evaluating the robustness of the predictions made by the trained classifier is essential. The model we built using XGBoost has been trained over a dataset comprising $\sim$ 47\% of signal events, the rest $\sim$ 53\% consisting solely of an irreducible background. The trained model appropriately identifies $\sim$81\% of the signal events ($\sim$38\% of the entire dataset) as signals, and $\sim$89\% of the background events ($\sim$47\% of the entire dataset) as background, achieving the precision of $\sim$ 85\%. The rest $\sim$19\% of the signal events have been misidentified as background, and $\sim$11\% of the background events have been misidentified as signal. 

\begin{figure}[!htb]
\centering 
\includegraphics[width=0.45\textwidth]{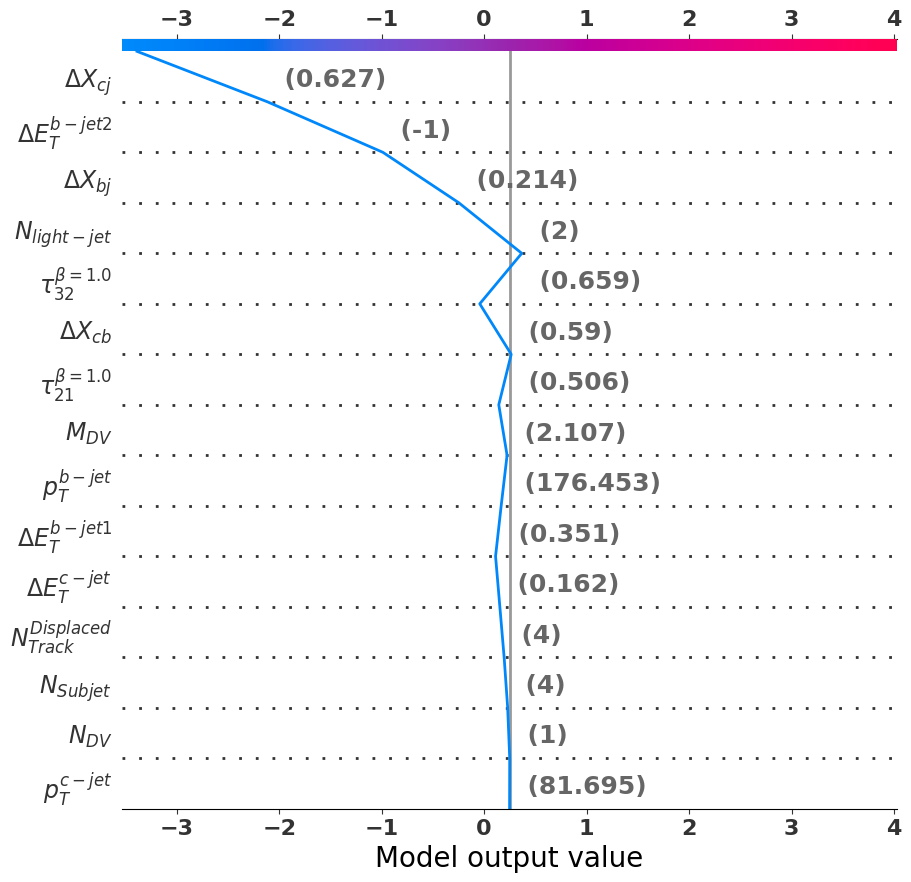}%decision_52_tree_clss.png
\hspace{1cm}
\includegraphics[width=0.45\textwidth]{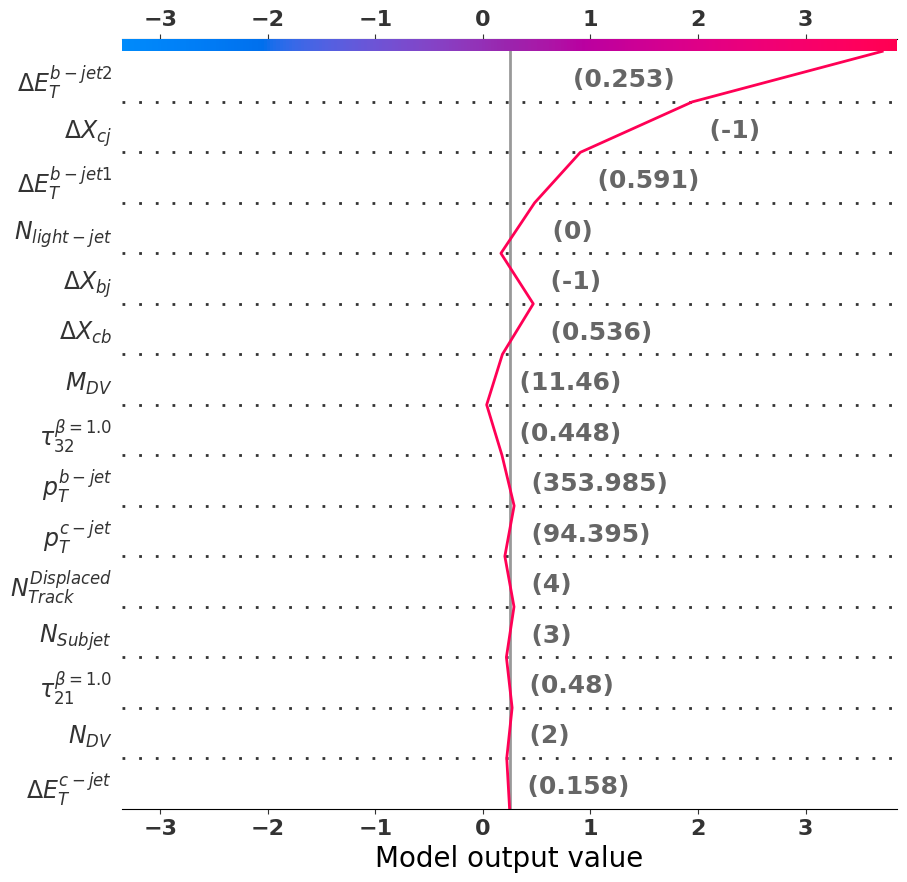}%decision_32_tree_clss.png
\caption{\label{Fig:13sp} The Shapley Decision Plot for a pair of properly identified events: \textit{Left:} Decision Plot for a background event. \textit{Right:} Decision Plot for a signal event. %Expected Value of the Model: 0.0147. The trajectories presented in the figures trace through the weighted Shapley Scores of the individual features, thereby portraying the contribution of each of them in classification of the concerned event.
}
\end{figure}

Below we display the decision making procedure for a pair of appropriately identified events (a signal event as a signal and a background as the background) by the model with the aid of the Shapley decision plot. Similar observations were also found for the misclassified events (a signal event wrongly classified as a background and vice versa). A Shapley decision plot for a particular event traces the trajectory of the cumulative Shapley score of individual features appearing in sequence. It commences from the base value (or the expected value/average prediction value) of the model and terminates at the predicted value for that particular event. If the final value is positive, the event is classified as a signal, otherwise as a background event. %The Shapley scores for each feature are expressed relative to the expected value of the model. Sum of the Shapley values of individual features, for a particular event, along with the expected value of the model yields the predicted value for that event as the model output. 
The trajectories presented in the figures are traced through the weighted Shapley scores of the individual features, thereby portraying the contribution of each of them in the classification of the concerned event. In Figure \ref{Fig:13sp}, a pair of correctly classified events, a signal as a signal and a background as a background event, are shown. 
We refrain from describing the effect of the individual features on the trajectory in detail, but it is noteworthy that the features chiefly responsible for proper identification of the signal (or background) event are those having positive (or negative) Shapley values globally. Note that since the Decision Plot portrays a local trait, the order of importance of the set of features can be different from that observed in the summary plot, which describes the model globally.

%We hope that the reader can make out the observations presented in Figure \ref{Fig:13sp} with ease.

%--------------------------------

\section{Summary and Outlook}  \label{Sec:Conc}

We focused on a BSM probe driven by FCNC-mediated rare top decay $t \to cH$ in the boosted regime, at a high luminosity run of the LHC, with a center-of-mass energy of 13 TeV. We considered the most abundant Higgs decay to $b\bar{b}$ for enhanced signal purity over backgrounds resembling similar event topologies. The selection of boosted regime minimizes the spurious SM backgrounds. 

We choose the production of highly boosted top quark in association with a W-boson. The boosted top would eventually manifest itself as a large-R collimated jet in the detector, encapsulating a 3-prong substructure, comprising one c-tagged jet and two b-tagged jets. The W boson would undergo leptonic decay, which provides the possibility of using the high leptonic $p_T$ and/or the high missing transverse energy as the trigger. We consider the possible Standard Model (SM) background processes that can mimic the signal under consideration: the most obvious one involves the same associate production, $pp → tW$, but the top quark undergoing the most abundant decay, $t \to bW$ followed by the hadronic decay of the W-boson, $W \to c\bar{s}$. The collimated large-R jet obtained in this case would also possess a three-prong substructure consisting of one b-tagged jet, one c-tagged jet and a light jet. This is the irreducible background for our analysis. The QCD backgrounds with robust cross sections, \textit{viz.}, $pp \to gW$ followed by the leptonic decay of the W boson, and $pp \to jj$ where $j$ refers to colored objects (quarks and gluons) with $p^{j}_{T, min}:$ 350 and 500 GeVs, can be minimized upon imposing the above mentioned trigger. 

For the effective reconstruction of the boosted top, manifested as a collimated large-R jet comprising 3-prong substructure: one c-tagged jet and two b-tagged jets with the b-tagged jet pairs appearing from the decay of a Higgs boson, we propose a new, dedicated top tagger. In parallel, we performed a comparative analysis of different kinematic features between the signal and the different backgrounds to comprehend and correlate their contrasting behavior for labeling each event as a signal or background. This enables the imposition of cuts for the optimization of signal events over existing backgrounds. We implement the Boosted Decision Tree(BDT) algorithm in our tagger, to identify signatures of rare top decay ($t \to cH$), as a probe of BSM Physics, from the most abundant decay mode of the top ($t \to bW$), along with QCD backgrounds. We compare the performance of our tagger with two predominantly used cut-based top taggers, namely the JHTopTagger and HEPTopTagger. Our top-tagger outperforms the existing top tagging algorithms, specifically for the above-mentioned cascade of rare-decay.

With spurious QCD Backgrounds eliminated upon the imposition of the trigger, we carry forward our analysis with the irreducible background and the signal processes to obtain an ROC contrasting the performance achieved by the Extreme Gradient Boosting (XGBoost) and the Adaptive Boosting (AdaBoost) algorithms. We observe that XGBoost shows better performance in properly identifying events than AdaBoost. Comparison between the efficiencies achieved with XGBoost and with the available ML-based toptaggers, namely DNN-based DeepTop, CNN-based taggers, and GNN-based LorentzNet shows that our model performs better in probing flavor-violating boosted top decays. We finally perform a Shapley analysis to assess the relative contributions of individual features towards event classifications. We retain the model built with XGBoost and implement the Tree Shapley Explainer. The Shapley analysis provides both global and local insights into the contribution of the model features towards the event classification.

Our motive behind the proposition and construction of such toptagger is to enhance efficient identification of $t \to cH$ decay mode, which if observed to deviate from the SM prediction, would inevitably support the BSM theories leading to relatively enhanced $t \to cH$ production. {Despite the commendable performance of our tagger over existing cut-based techniques and more advanced Deep-Learning-driven approaches to detect FCNC-mediated top-quark decays through jet-substructure techniques, the proposed method yields a lower signal significance in terms of the limits on $t \to cH$ branching ratio, compared to the existing limits from the LHC run-2 data. Although it is inadequate to confirm any evidence or discovery, it is substantial to redirect future collider searches for rare top decays.} Besides, the approach taken for our analysis is generic and model-independent. Similar prescriptions can be followed for the fabrication of newer algorithms for probing BSM Physics from boosted large-R jets with substructures. There are various BSM theories which may lead to topologies similar to those we considered for our analysis. One such example can be the SM augmented with the vector-like quarks, which may manifest a collider signature of similar kind. Hence, this work, by far, has potential prospects and would inevitably lead to more improvised, sophisticated jet-tagging algorithms for BSM probes with enhanced performances. %With the high-luminosity runs of the LHC in action and few advanced particle colliders in the pipeline, these algorithms would be pivotal in disfavoring and reinforcing extensions proposed over the SM.

%%%%%%%%%%%%%%%%%%%%%%%%%%%%%%%%%%%%%%%%%%%%%%%%%%%%%%%%%%%%%%
\section{Acknowledgments}
%=============================================================================================

The work of AC is funded by the Department of Science and Technology, Government of India, under Grant No. IFA18-PH 224 (INSPIRE Faculty Award). SD acknowledges the SERB Core Grant (CRG/2018/004889) for financial support. The authors also acknowledge fruitful discussions with Rameswar Sahu and Kirtiman Ghosh from IOP, Bhubaneswar.

%%%%%%%%%%%%%%%%%%%%%%%%%%%%%%%%%%%%%%%%%%%%%
%%%% Appendix starts here 
%%%%%%%%%%%%%%%%%%%%%%%%%%%%%%%%%%%%%%%%%%%%%

\appendix

%--------------------------------------------------------------
\section{Event Simulation and Reconstruction Details}
\label{Sec:App1}
%--------------------------------------------------------------

We follow a model-independent generic framework in the rare top decay channel $t \to cH$, and use the {\tt FeynRules} \cite{feynrules1, feynrules2} model files implemented by the authors of \cite{Buchkremer:2013bha}. The UFO output thus obtained is provided to {\tt MadGraph} (version 3.4.1) \cite{Alwall:2011uj} to generate hard scattering events. The desired decays are obtained using MadSpin. The center of mass energy of the run card is set to be 13 TeV, and the renormalization and factorization scales are set to be $H_T$/2, where $H_T$ denotes the sum of the transverse energy of each parton. We choose
{\tt NNPDF23-lo-as-0130-qed} \cite{NNPDF:2014otw} as the Parton Distribution Function (PDF). The requirement of boosted resonant production of the top quark has been met upon setting the minimum transverse momentum ($p_T$) of the top, at the particle level, to be 350 GeV. The LHEF output of the events thus generated are then plugged in {\tt PYTHIA8} (version 8.310) \cite{Bierlich:2022pfr} to account for the initial- and final-state radiations, parton showering and hadronization. The HepMC \cite{hepmc} output obtained by PYTHIA8 is then passed to {\tt Delphes} (version 3.5.0) \cite{delphes} to perform a fast detector simulation, using the CMS card available within Delphes.

%==========================================
\section{Jets and their substructures, and flavor tagging}
\label{Sec:App2}

We reconstruct jets using {\tt FastJet} \cite{fastjet1, fastjet2} with the anti-kT jet clustering algorithm \cite{antikt}, setting the jet radius parameter $ R = 1.0$. Note that the choice of the large-R jet radius was made by a detailed analysis using truth-level event information. For example, we calculate the fraction of events where either the leading large-R jet includes all the decay products of the top quark, or the direction of the top quark is along the large-R jet axis. We observe that R = 1.0 is an optimal choice for the region of interest, as a higher value of R would include more contamination from underlying events (UE) and multiparticle interactions (MPI) \footnote{Along with the fixed-R jet algorithm, one may use variable-R \cite{Krohn:2009zg} or the dynamic radius algorithm \cite{Mukhopadhyaya:2023rsb} to follow a more robust jet clustering algorithm.} To reduce contamination from the UE / MPIs mentioned above, we apply the jet trimming technique \cite{Krohn:2009th} where the $k_T$ algorithm is used with the subjet radius 0.3 and the $p_T$ fraction of subjets over the large-R jet is chosen to be 0.03.    

%---------------------------------------------
%---------------------------------------------

\begin{figure}[!htb]
\centering  
{\includegraphics[width=0.4\textwidth]{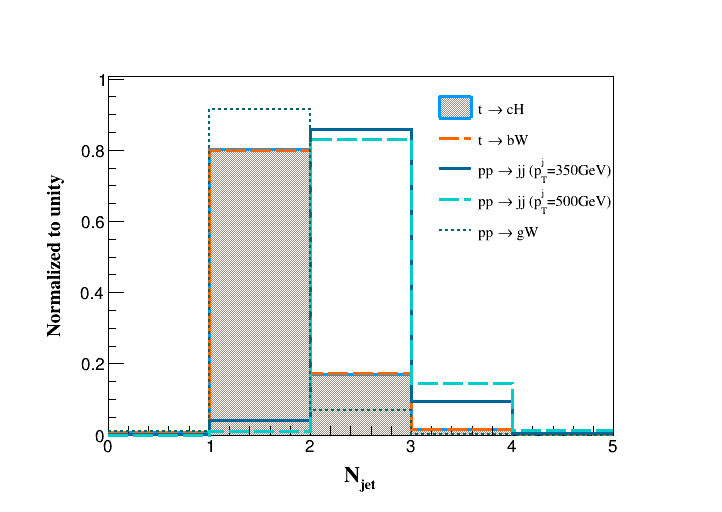}  }
\hfill
\caption{\label{Fig:7} Distributions of large-R jet multiplicity for the signal and SM Background events.}  
\end{figure}

In Figure \ref{Fig:7}, we present the multiplicity of large-R jets for the signal and all possible SM background events. The multiplicity of jets for both signal events (where the top decays through the $cH$ mode) and the irreducible background (where the top decays through the dominant mode $bW$) has been observed to be dominant around unity. The reason behind this can be traced to the fact of production of a single, boosted top quark from the $pp$ collision along with a leptonically decaying $W$-boson, and the formation of highly collimated stable decay products of top clustered in a large-R jet. Because of the similar kinematics of the top quark for both the signal and the irreducible background, their respective jet multiplicities are statistically identical. As one notices, the single large-R jet events for the QCD events are much smaller than those of the signal and the irreducible background, except the QCD background comprising $pp \to g W$ events. We re-emphasize at this point that the QCD background can be minimized with the trigger discussed in Section \ref{Sec:EvSim}.

\begin{figure}[!htb]
\centering   % \begin{center}/ \end{center} takes some additional vertical space
\mbox{
{\includegraphics[width=.4\textwidth]{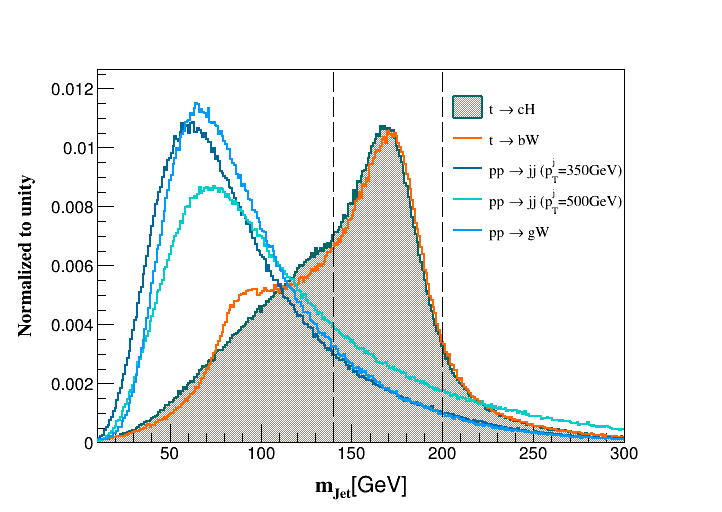}  }
\hfill
{\includegraphics[width=.4\textwidth]{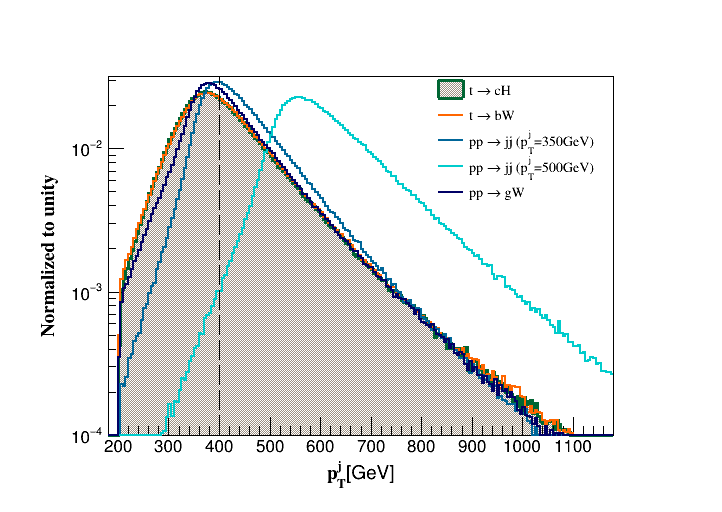} }
}
	\caption{\label{Fig:8} Distribution of the mass (left) and the $p_{T}$ (right) of the leading large-R jet for signal and all the background events. }  
\end{figure}

The next observation, presented in Figure \ref{Fig:8}, shows the kinematics of the large-R jet. These jets, reconstructed for each event after the imposition of detector effects, are arranged in descending order of their respective $p_T$s with a minimum $p_T$ of 200 GeV. 
The left plot indicates the mass of the leading large-R jet peaks around 172 GeV for signal and irreducible background, while for spurious backgrounds it's peak between 50 and 70 GeV with an exponential decline. The $p_T$ distribution of the leading jet shows similar traits for the signal and irreducible background events and peaks at $\sim$380 GeV with a skew towards the right. In order to minimize the contribution of spurious backgrounds, we set the large-R jet-$p_T$ cut of 400 GeV for all our future analysis thereby preserving $\sim$75 \% of the signal and the irreducible background events. %Having analyzed different aspects of the large-R jet, we now continue our discussion for the rest of this section focusing on the substructures of the jet and their varied traits, considering it as a candidate jet for the boosted top quark.

%%%%%%%%%%%%%%%%%%%%%%%%%%%
\vspace{0.1cm}
$\bullet$ ~{\bf Substructures inside the top-candidate jet and their flavour tagging:} 

\begin{figure}[!htb]
\centering   % \begin{center}/ \end{center} takes some additional vertical space
{\includegraphics[width=.4\textwidth]{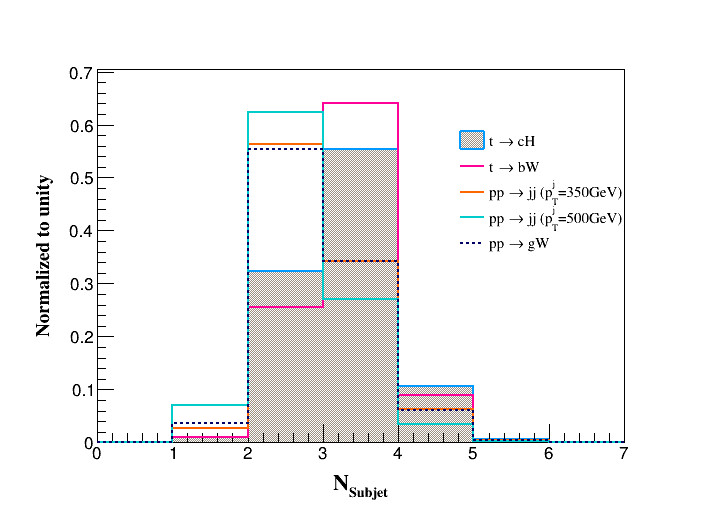}  }
\caption{\label{Fig:15}Normalized distribution of subjet multiplicities, for signal, irreducible background and the QCD events.}
\end{figure}

With our understanding of the internal structure of the large-R jet so far, we now proceed to form the subjets using the constituents of the boosted large-R jet. So, we recluster the constituents with subjet radius 0.3 and a minimum $p_{T}$ of 20 GeV. As anticipated from the observations made with Figure \ref{Fig:14}, the signal and the irreducible background have subjet multiplicities (see Figure \ref{Fig:15}) predominantly peaking at 3, in contrast to that for the QCD events with peaks at 2. The contrasting behavior of $\tau_{32}$ between the signal and the irreducible background asserts the higher relative abundance of irreducible background events with a subjet multiplicity of 3, over the signal, the reason being again the mass asymmetry. 

\begin{figure}[!htb]
\centering   % \begin{center}/ \end{center} takes some additional vertical space
\mbox{
{\includegraphics[width=.3\textwidth]{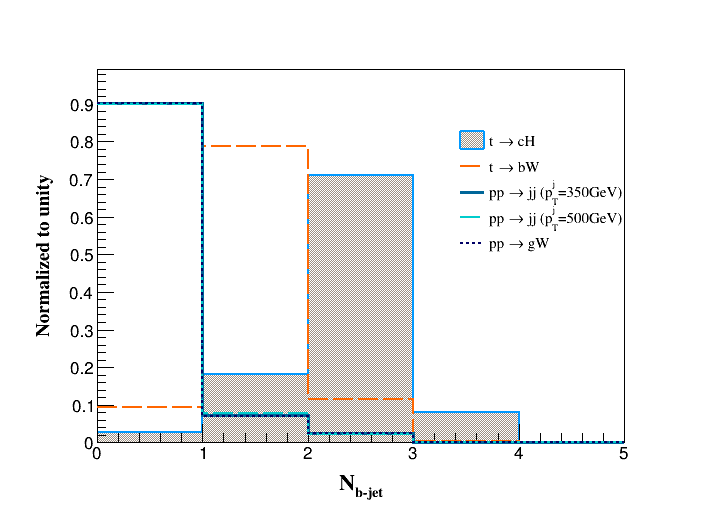}  }
{\includegraphics[width=.3\textwidth]{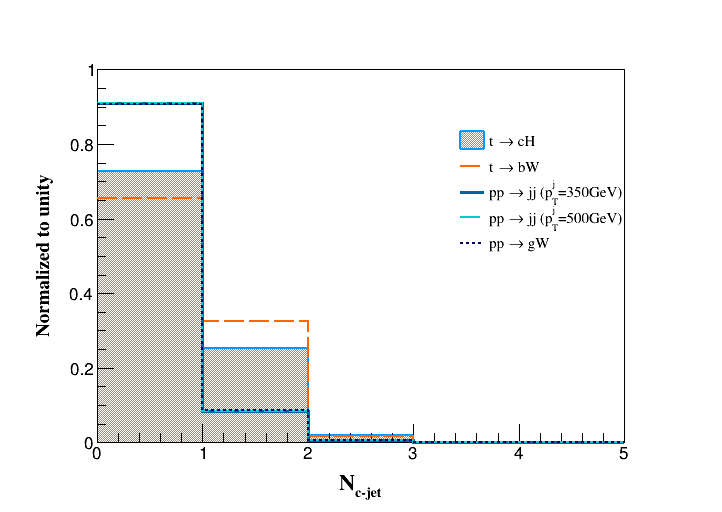}  }
{\includegraphics[width=.3\textwidth]{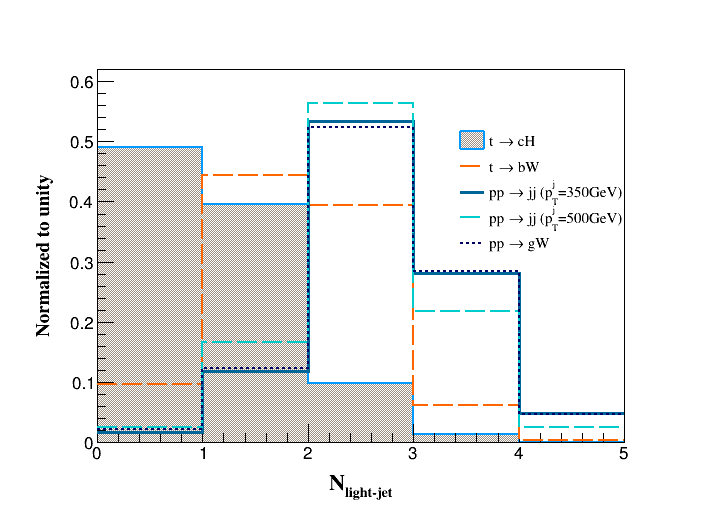}  }}
\caption{\label{Fig:16} Normalized distribution of the multiplicities of b-tagged jet, c-tagged jet and the light-jet, for the signal, background and QCD processes.}
\end{figure}

Moving forward, to tag a subjet with a specific flavor, b or c, we consider the b- and c-hadrons that appeared within the radius of the large-R jet formed after the hadronization of the b- and c-quarks. If the angular separation $(\Delta R)$ of a hadron of certain flavor (b or c) from a given subjet axis is less than the subjet radius, then the concerned subjet is tagged as the jet of that very flavor. The subjets, not tagged with any of the b- or c-hadrons, are labeled as light jets. For the identification of b- and c-tagged jets, we also consider detector effects by incorporating b-tagging and c-tagging efficiencies at 70\% and 40\% respectively, while for their mistagging rates we follow a $p_T$-dependent formula provided in \cite{ATLAS:2017bcq,ATLAS:2015prs}. Note that b-tagging precedes c-tagging to avoid overlap. Once a subjet is tagged with a flavor, it is excluded from further tagging. Figure \ref{Fig:16} illustrates the multiplicities of subjets tagged with flavor for the signal and different SM backgrounds, following this prescription.

In Figure \ref{Fig:16} (left), the multiplicity of b-tagged jets peaks at two for the signal (2b-tagged jets from Higgs boson decay), one for the irreducible background (single b-tagged jet from top decay), and zero for QCD events. The multiplicity of the c-tagged jet (Figure \ref{Fig:16} (center)) in the signal, originating from c-quarks in top decay, is found to be insufficiently incorporated within the large-R jet. The asymmetry involved in top decay to the Higgs boson leaves less available phase space for the c-quarks to obtain adequate energy, and in the substantial angular separation between the $b \bar{b}$ pair from the Higgs decay, which collectively leads the large-R jet abundantly containing the hadronised products of the b-quarks and less frequent incorporation of c-hadrons. The irreducible background follows a trend similar to that of the signal. The light-jet multiplicity in Figure \ref{Fig:16} (right) peaks at zero for the signal, with occasional instances of one light jet due to the QCD radiations from the heavier partons and heavy flavour mistagging. Conversely, the irreducible background peaks at one due to s-quarks from W-decay and the QCD backgrounds peak at three due to the scarcity of heavy-quark flavors at the partonic level.

%--------------------------------------------------------------
\section{Reconstruction of Displaced Tracks and Displaced Vertices }
\label{Sec:App3}
\begin{figure}[!htb]
\centering   
\mbox{
{\includegraphics[width=.4\textwidth]{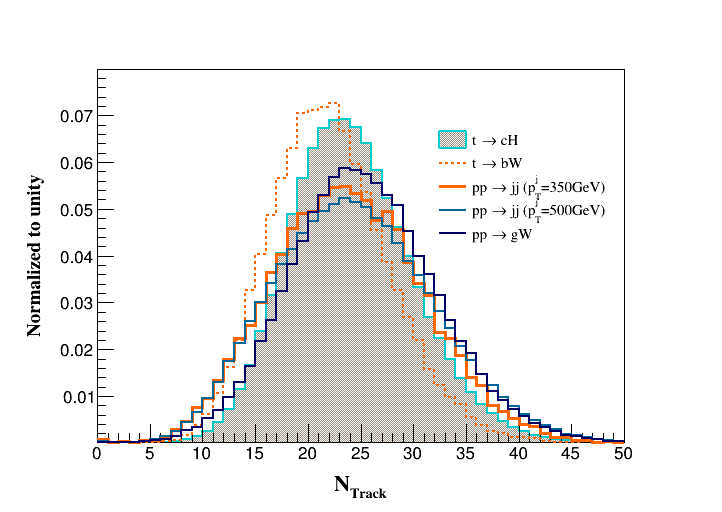}  }
\hfill
{\includegraphics[width=.4\textwidth]{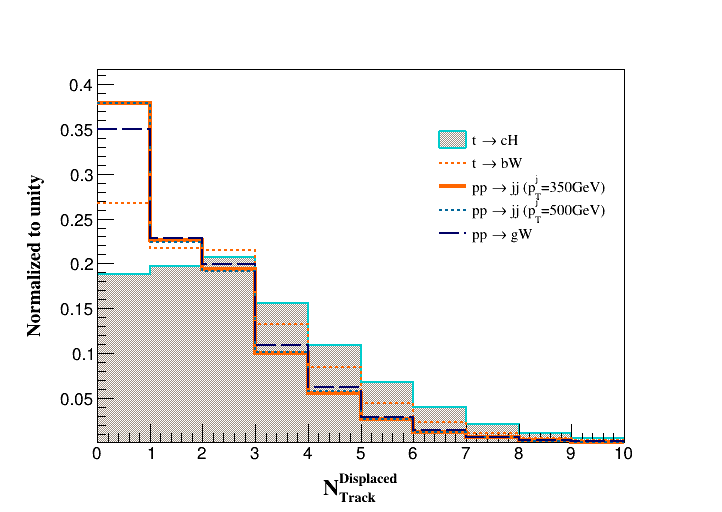} }
}
\caption{\label{Fig:11}\textit{Left}: Normalized distribution of total number of charged tracks within the large-R jet. \textit{Right}: Normalized distribution of the displaced tracks out of the registered tracks in the large-R jet. The plots show a comparative analysis between the signal, irreducible background and the QCD events.}
\end{figure}

%--------------------------------------------------------------
A jet, as perceived in the detector to be a localized cluster of heavy-energy deposits, consists of different fundamental and composite particles, electromagnetically charged and neutral. The charged particles leave their respective traces on the silicon trackers surrounding the interaction point in the detector. The uncharged constituents are directly detected with their energy deposition at the electromagnetic calorimeter or at the hadronic calorimeter. The quarks originating from the top decay (and associated radiative effects, lead to large number of charged hadronic bound states, and thereby many tracks, as seen in the left panel of Figure \ref{Fig:11}. Those tracks with $p_{T}$s greater than 2 GeV are considered here. 

%------------------------------
%---------------------------------------

Even though the track multiplicities are large for both signal and background events, presence of more heavy quarks for the signal events should lead to more displaced energetic tracks. 

In order to label a track to be displaced, we impose an additional cut on the perpendicular component of the impact parameter of the given track, $\lvert d_{\perp} \rvert > 1$ with a resolution uncertainty ($\sigma_{d_{\perp}}$) > 4; the choices follow Ref.\cite{CMS:2018tuo,ATLAS:2019fwx,ATLAS:2019kpx}. In Figure \ref{Fig:11} (right) we show the multiplicity of displaced tracks inside the top-candidate jet for the signal and the backgrounds. As anticipated, the signal predominantly containing two b- and one c-hadrons leads to more events with three displaced tracks with respect to the irreducible background, while the irreducible background with one b- and one c-hadrons has most events with number of displaced tracks between one and two. The displaced track multiplicity for the QCD processes is substantially less. 

\begin{figure}[!htb]
\centering   
{\includegraphics[width=.4\textwidth]{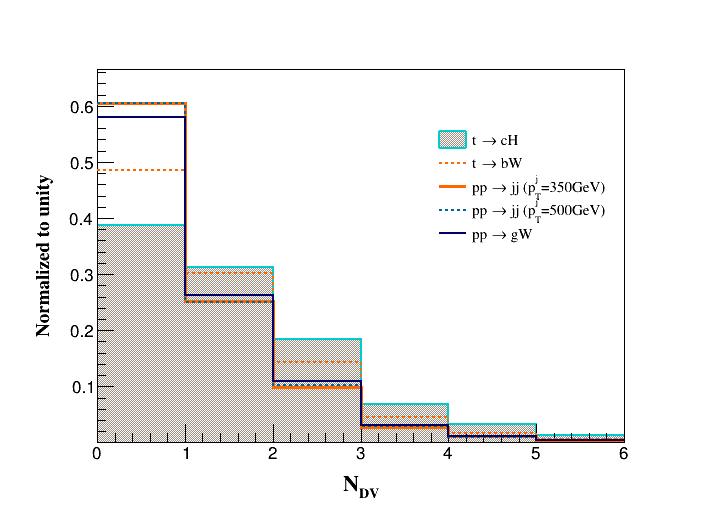}  }
{\includegraphics[width=.4\textwidth]{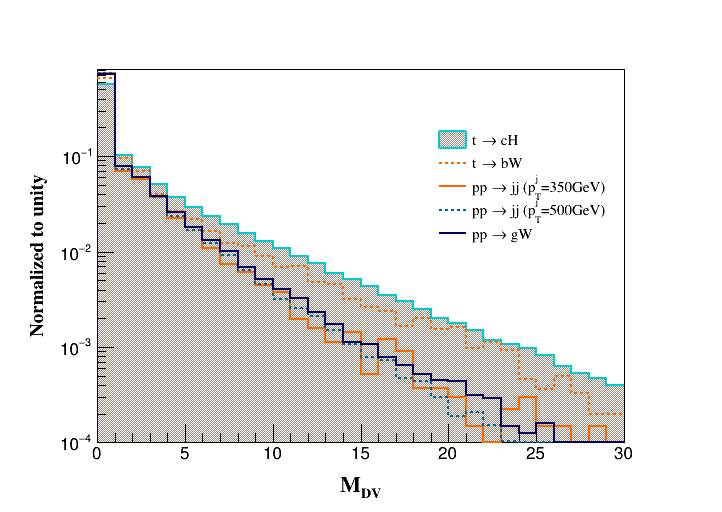}  }
\caption{\label{Fig:12}\textit{Left}: Normalized distribution of the net number of vertices formed by the displaced tracks. \textit{Right}: Normalized distribution of the invariant mass reconstructed from the daughter tracks. The plot provides a comparative analysis between the signal, irreducible background and the QCD events.}
\end{figure}

The retrieval of displaced tracks leads to the reconstruction of the displaced vertex. The silicon trackers provide the initial information about these tracks. Following \cite{Chakraborty:2020cpa}, we choose those tracks for which the effective displacement satisfies ,
\beq
\lvert \Delta X_{ij} \rvert < 10^{-3} ~\text{GeV}^{-1}, \quad \lvert \Delta Y_{ij} \rvert < 10^{-3} ~\text{GeV}^{-1}, \quad \lvert \Delta Z_{ij} \rvert < 10^{-3} ~\text{GeV}^{-1},
\eeq
where $j$ and $i$ denote respectively the point where one track vanishes and the other track(s) emerges. The invariant mass of the vertex ($j^{th}$ index) has been then calculated from the collective 4-momentum of the corresponding re-emerging tracks,
\beq
M_{DV}^2 ~=~ \left( \sum_i E_{i} \right)^{2}  - \left( \sum_i \vec{p_{i}} \right)^{2},
\eeq
where $i$ runs over all the tracks associated to the vertex. Figure \ref{Fig:12} represents the multiplicity of displaced vertices and the reconstructed mass of the displaced vertices calculated using the displaced tracks of the signal and background events.

With the reasons discussed previously, the multiplicity for larger number of displaced tracks is higher for the signal than that of the different SM backgrounds, and so the behavior observed in the multiplicity and reconstructed mass of the reconstructed displaced vertices are expected. 
%%%%%%%%%%%%%%%%%%%%%%%%%%%%%%%%%%%%%%%%%%%%%%%%

%%%%%%%%%%%%%%%%%%%%%%%%%%%%%%%%%%%%%%%%%%%%%%%%%%%%%%
%%%%%%%%

\section{Dependence plot among the input features in SHAP}
\label{Sec:App-dependence}

\begin{figure}[!htb]
%\centering 
\includegraphics[width=0.450\textwidth]{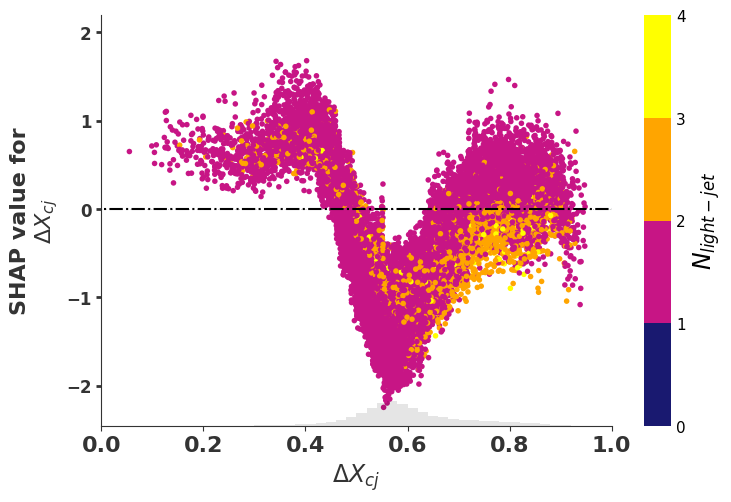}%delX_cj-nljet.png
\hspace{1cm}
\includegraphics[width=0.450\textwidth]{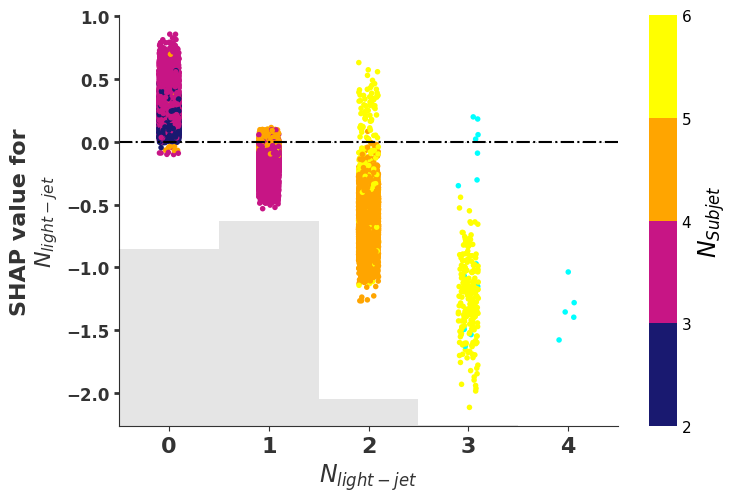}%nljet-nsjet.png
\caption{\label{Fig:3sp} Left: A scatter plot between ${\rm \Delta X_{cj}}$ and the respective Shapley value, portraying the behavior in connection with the light jet multiplicity. The y-axis of the figure represents the Shapley value of ${\rm \Delta X_{cj}}$, which is plotted along the x-axis. Each point on the figure represents an event. The color coding speaks of the light jet multiplicity($N_{light-jet}$) for each event. Right: A bar-chart of the light jet multiplicity, with the event-by-event Shapley score associated with it presented along y-direction. The color code represents subjet multiplicity ($N_{Subjet}$) for each event under consideration. The gray shaded histogram gives the distribution of the lightjet multiplicities.
}
\end{figure}

We next discuss the effect of one feature over another for the classifier, considering all the events that include the signal and the irreducible background for their generation. A SHAP dependence plot is a visualization tool that illustrates the relationship between a feature and the prediction of the model. It shows how the model prediction varies as the value of the feature changes. To uncover potential interaction effects between the features, we can modify the dependence plot by adding color coding based on a second feature. Thus, we determine the Shapley score for the concerned feature for each event and color coded the data points (referring to the events) by the values assumed by the other feature. We refrain from quantifying and presenting the contribution of the pairwise interaction of the features to the performance of the classifier. The Shapley score for this case is straightforward and is given by
\beq
\displaystyle \varphi _{i,j}(v) ~=~ {\frac {|S|!\;(n-|S|-2)!}{(n-1)!}} ~ \sum _{\substack{S\subseteq N \setminus \{i,j\}} \\ i \neq j} ~ \chi_{ij},
\eeq
where, 
\[
\chi_{ij} = \left( v(S \cup \{i,j\}) ~-~ \left( v(S \cup \{i\}) + v(S \cup \{j\}) + v(S) \right) \right),
\]
and $i, ~j$ index the feature pairs.
%%%%%%%%%%%%%%%%%%%%%%%%%%%%%%%%%%%%%%%%%%%%%%%%%%%%%
Our quest continues with the study of the behavior of ${\rm \Delta X_{cj}}$ in association with the multiplicity of light jets inside the large-R  jet. The left panel of Figure \ref{Fig:3sp}  presents a scatter plot between the value of ${\rm \Delta X_{cj}}$ for each event and the respective Shapley score for that feature value. 
%As for the Figure \ref{Fig:1sp} (left), both signal and irreducible background events have been taken into account.

At the bottom of Figure \ref{Fig:3sp} (left), a gray histogram presents the distribution of ${\rm \Delta X_{cj}}$ for all events, including the signal and the irreducible background. The histogram of ${\rm \Delta X_{cj}}$ peaks at the value of $~\sim 0.6$, precisely where the scattered plot between ${\rm \Delta X_{cj}}$ and the respective Shapley value for each event achieves the minima. Since the points above the $y ~=~ 0$ axis are labeled as signal events by the classifier, this behavior around ${\rm \Delta X_{cj}} ~\sim 0.6$ indicates that the said region of the parameter space is background dominated. Let us remind ourselves that the background contains an s-jet predominantly identified as the light jet, and hence the residual invariant mass fraction of the light jet and the c-tagged jet predominantly lies around the resonance mass of the $W$-boson ($\sim$80 GeV). Assuming the mass of the large-R  jet to be around the resonance peak of the top quark ($\sim$172 GeV), ${\rm \Delta X_{cj}}$ for the background peaks is around 0.6. Since the light jet for the signal originates principally from the radiative effects of QCD, ${\rm \Delta X_{cj}}$ for the signal, is randomly distributed along the feature space. As the figure appropriately portrays, the data points distant from the peak around ${\rm \Delta X_{cj}} ~\sim 0.6$ tend to be signal-like. As anticipated, every event possesses at least one light jet (color coded magenta), with a minute fraction of events having higher light jet multiplicity, present in both the predicted classes (signal and background), owing to the radiative effects.
%%%%%%%%%%%%%%%%%%%%%%%%%%%%%%%%%%%%%%%%%%%%

%\begin{figure}[!htb]
%\centering 
%\includegraphics[width=0.60\textwidth]{figures/nljet-nsjet.png}
%\caption{\label{Fig:4sp} A bar-chart of the light jet multiplicity, with the event-by-event Shapley score associated with it presented along y-direction. The color code represents subjet multiplicity ($N_{Subjet}$) for each event under consideration. The instances with subjet multiplicities greater than five are marked with color cyan. The gray shaded histogram gives the distribution of the lightjet multiplicities.}
%\end{figure}

We continue our analysis with light jet multiplicities for a given event, in association with the multiplicity of subjet for the same event. Figure \ref{Fig:3sp} (right) shows a bar diagram for different light jet multiplicities, with the vertical axis referring to the Shapley score for each data point corresponding to an event classified according to the respective light jet multiplicity, and color coded to carry the information of the subjet multiplicity for the same event. The light jet multiplicity distribution displayed by the gray shaded histogram in the Figure \ref{Fig:3sp} (right) and the corresponding positive Shapley score reinforces the fact that the classifier predominantly predicts events with no light jet as the signal and higher light jet multiplicities as background.  %This is in accordance to the reasoning we made before, as in parton level the large-R jet for the signal does not contain any light quark, and the light jet only appears from the mistagging or the radiative effects. These events are characterized by subjet mutipliciities chiefly ranging between two and three, although few events can be observed having higher number of subjets followed from QCD radiative effects. The events with higher light jet multiplicities are mainly predicted as background by the model. 

In our analysis, we study the interaction effects of other relevant features also. We conclude here that the behavior of different pairs of features in determining signals and background is well understood and addressed with the underlying physics. In fact, globally what follows as an inevitable explanation to some of the observed discrepancies are actually coming due to the interference/interaction effects. The overall prediction of the classifier model to segregate events into signal and background is the combined effects of all these features.
\bibliographystyle{unsrt}
%\bibliographystyle{abbrvnat}
%\bibliographystyle{IEEEtran}%IEEEtran,plainnat
%\setcitestyle{ maxcitenames=3}
%\bibliography{reference}

\begin{thebibliography}{10}

\bibitem{Glashow:1970gm}
S.~L. Glashow, J.~Iliopoulos, and L.~Maiani.
\newblock {Weak Interactions with Lepton-Hadron Symmetry}.
\newblock {\em Phys. Rev. D}, 2:1285--1292, 1970.

\bibitem{article:tawj}
A.~Tumasyan, W.~Adam, J.~Andrejkovic, Thomas Bergauer, S.~Chatterjee,
  M.~Dragicevic, A.~Valle, Rudolf Frühwirth, M.~Jeitler, N.~Krammer, Luzian
  Lechner, D.~Liko, I.~Mikulec, Peter Paulitsch, F.~Pitters, J.~Schieck,
  Songyou Xie, M.~Spanring, S.~Templ, and W.~Vetens.
\newblock {Measurement of the top quark mass using events with a single
  reconstructed top quark in pp collisions at $\sqrt{s}$ = 13 TeV}.
\newblock {\em Journal of High Energy Physics}, 2021, 12 2021.

\bibitem{Quadt2007TopQP}
Arnulf Quadt.
\newblock {Top quark physics at hadron colliders}.
\newblock {\em Eur. Phys. J. C}, 48:835--1000, 2006.

\bibitem{aubert2012particle}
J.J. Aubert, R.~Gastmans, and J.M. G{\'e}rard.
\newblock {\em Particle Physics: Ideas and Recent Developments}.
\newblock Nato Science Series C:. Springer Netherlands, 2012.

\bibitem{fcncBr1}
J.~A. Aguilar-Saavedra.
\newblock {Top flavor-changing neutral interactions: Theoretical expectations
  and experimental detection}.
\newblock {\em Acta Phys. Polon. B}, 35:2695--2710, 2004.

\bibitem{fcncBr2}
Gauhar Abbas, Alejandro Celis, Xin-Qiang Li, Jie Lu, and Antonio Pich.
\newblock {Top FCNC decays in the aligned two-Higgs doublet model}.
\newblock {\em PoS}, EPS-HEP2015:326, 2015.

\bibitem{fcncBr3}
Shyam Balaji.
\newblock $cp$ asymmetries in the rare top decays
  $t\ensuremath{\rightarrow}c\ensuremath{\gamma}$ and
  $t\ensuremath{\rightarrow}cg$.
\newblock {\em Phys. Rev. D}, 102:113010, Dec 2020.

\bibitem{Kiers:2011sv}
Ken Kiers, Tal Knighton, David London, Matthew Russell, Alejandro Szynkman, and
  Kari Webster.
\newblock {Using t -\ensuremath{>} b \textbackslash{}bar{b} c to Search for New
  Physics}.
\newblock {\em Phys. Rev. D}, 84:074018, 2011.

\bibitem{Bardhan:2016txk}
Debjyoti Bardhan, Gautam Bhattacharyya, Diptimoy Ghosh, Monalisa Patra, and
  Sreerup Raychaudhuri.
\newblock {Detailed analysis of flavor-changing decays of top quarks as a probe
  of new physics at the LHC}.
\newblock {\em Phys. Rev. D}, 94(1):015026, 2016.

\bibitem{Chen:2023eof}
Chuan-Hung Chen, Cheng-Wei Chiang, and Chun-Wei Su.
\newblock {Top-quark FCNC decays, LFVs, lepton $g-2$, and $W$ mass anomaly with
  inert charged Higgses}.
\newblock 1 2023.

\bibitem{Altmannshofer:2019ogm}
Wolfgang Altmannshofer, Brian Maddock, and Douglas Tuckler.
\newblock {Rare Top Decays as Probes of Flavorful Higgs Bosons}.
\newblock {\em Phys. Rev. D}, 100(1):015003, 2019.

\bibitem{Hung:2017tts}
P.~Q. Hung, Yu-Xiang Lin, Chrisna~Setyo Nugroho, and Tzu-Chiang Yuan.
\newblock {Top Quark Rare Decays via Loop-Induced FCNC Interactions in Extended
  Mirror Fermion Model}.
\newblock {\em Nucl. Phys. B}, 927:166--183, 2018.

\bibitem{Larios:2006pb}
F.~Larios, R.~Martinez, and M.~A. Perez.
\newblock {New physics effects in the flavor-changing neutral couplings of the
  top quark}.
\newblock {\em Int. J. Mod. Phys. A}, 21:3473--3494, 2006.

\bibitem{Banerjee:2018fsx}
Shankha Banerjee, Mikael Chala, and Michael Spannowsky.
\newblock {Top quark FCNCs in extended Higgs sectors}.
\newblock {\em Eur. Phys. J. C}, 78(8):683, 2018.

\bibitem{Kim:2015zla}
C.~S. Kim, Yeo~Woong Yoon, and Xing-Bo Yuan.
\newblock {Exploring top quark FCNC within 2HDM type III in association with
  flavor physics}.
\newblock {\em JHEP}, 12:038, 2015.

\bibitem{ATLAS:2022gzn}
Georges Aad et~al.
\newblock {Search for flavour-changing neutral current interactions of the top
  quark and the Higgs boson in events with a pair of \ensuremath{\tau}-leptons
  in pp collisions at $ \sqrt{s} $ = 13 TeV with the ATLAS detector}.
\newblock {\em JHEP}, 2306:155, 2023.

\bibitem{CMS:2021hug}
A.~Tumasyan et~al.
\newblock {Search for Flavor-Changing Neutral Current Interactions of the Top
  Quark and Higgs Boson in Final States with Two Photons in Proton-Proton
  Collisions at $\sqrt{s}=13\text{ }\text{ }\mathrm{TeV}$}.
\newblock {\em Phys. Rev. Lett.}, 129(3):032001, 2022.

\bibitem{CMS:2021gfa}
Armen Tumasyan et~al.
\newblock {Search for flavor-changing neutral current interactions of the top
  quark and the Higgs boson decaying to a bottom quark-antiquark pair at $
  \sqrt{s} $ = 13 TeV}.
\newblock {\em JHEP}, 02:169, 2022.

\bibitem{CMS:2017bhz}
Albert~M Sirunyan et~al.
\newblock {Search for the flavor-changing neutral current interactions of the
  top quark and the Higgs boson which decays into a pair of b quarks at
  $\sqrt{s}=$ 13 TeV}.
\newblock {\em JHEP}, 06:102, 2018.

\bibitem{Greljo:2014dka}
Admir Greljo, Jernej~F. Kamenik, and Joachim Kopp.
\newblock {Disentangling Flavor Violation in the Top-Higgs Sector at the LHC}.
\newblock {\em JHEP}, 07:046, 2014.

\bibitem{CMS:Wcs}
Armen Tumasyan et~al.
\newblock {Precision measurement of the W boson decay branching fractions in
  proton-proton collisions at 13 TeV}.
\newblock {\em Phys. Rev. D}, 105(7):072008, 2022.

\bibitem{Chakraborty:2023dhw}
Amit Chakraborty, Amandip De, Rohini~M. Godbole, and Monoranjan Guchait.
\newblock {Tagging a Boosted Top quark with a $\tau$ final state}.
\newblock 4 2023.

\bibitem{JHToptagger}
David~E. Kaplan, Keith Rehermann, Matthew~D. Schwartz, and Brock Tweedie.
\newblock Top tagging: A method for identifying boosted hadronically decaying
  top quarks.
\newblock {\em Phys. Rev. Lett.}, 101:142001, Oct 2008.

\bibitem{Plehn:2011tg}
Tilman Plehn and Michael Spannowsky.
\newblock {Top Tagging}.
\newblock {\em J. Phys. G}, 39:083001, 2012.

\bibitem{Thaler:2008jutagger}
Jesse Thaler and Lian-Tao Wang.
\newblock {Strategies to Identify Boosted Tops}.
\newblock {\em JHEP}, 07:092, 2008.

\bibitem{Kasieczka:2017nvn}
Gregor Kasieczka, Tilman Plehn, Michael Russell, and Torben Schell.
\newblock {Deep-learning Top Taggers or The End of QCD?}
\newblock {\em JHEP}, 05:006, 2017.

\bibitem{Butter:2017cot}
Anja Butter, Gregor Kasieczka, Tilman Plehn, and Michael Russell.
\newblock {Deep-learned Top Tagging with a Lorentz Layer}.
\newblock {\em SciPost Phys.}, 5(3):028, 2018.

\bibitem{Macaluso:2018tck}
Sebastian Macaluso and David Shih.
\newblock {Pulling Out All the Tops with Computer Vision and Deep Learning}.
\newblock {\em JHEP}, 10:121, 2018.

\bibitem{Kasieczka:2019dbj}
Anja Butter et~al.
\newblock {The Machine Learning landscape of top taggers}.
\newblock {\em SciPost Phys.}, 7:014, 2019.

\bibitem{Dreyer:2020brq}
Fr\'ed\'eric~A. Dreyer and Huilin Qu.
\newblock {Jet tagging in the Lund plane with graph networks}.
\newblock {\em JHEP}, 03:052, 2021.

\bibitem{Andrews:2021ejw}
Michael Andrews et~al.
\newblock {End-to-end jet classification of boosted top quarks with the CMS
  open data}.
\newblock {\em EPJ Web Conf.}, 251:04030, 2021.

\bibitem{Chakraborty:2020yfc}
Amit Chakraborty, Sung~Hak Lim, Mihoko~M. Nojiri, and Michihisa Takeuchi.
\newblock {Neural Network-based Top Tagger with Two-Point Energy Correlations
  and Geometry of Soft Emissions}.
\newblock {\em JHEP}, 07:111, 2020.

\bibitem{Alvarez:2022qoz}
Ezequiel Alvarez, Manuel Szewc, Alejandro Szynkman, Santiago~A. Tanco, and
  Tatiana Tarutina.
\newblock {Exploring unsupervised top tagging using Bayesian inference}.
\newblock {\em SciPost Phys. Core}, 6:046, 2023.

\bibitem{Bhattacherjee:2022gjq}
Biplob Bhattacherjee, Camellia Bose, Amit Chakraborty, and Rhitaja Sengupta.
\newblock {Boosted top tagging and its interpretation using Shapley values}.
\newblock 12 2022.

\bibitem{ATLAS:2021wkg}
{Identification of hadronically-decaying top quarks using UFO jets with ATLAS
  in Run 2}.
\newblock 2021.
\newblock All figures including auxiliary figures are available at
  https://atlas.web.cern.ch/Atlas/GROUPS/PHYSICS/PUBNOTES/ATL-PHYS-PUB-2021-028.

\bibitem{ATLAS:2020szu}
{Boosted hadronic vector boson and top quark tagging with ATLAS using Run 2
  data}.
\newblock 2020.
\newblock All figures including auxiliary figures are available at
  https://atlas.web.cern.ch/Atlas/GROUPS/PHYSICS/PUBNOTES/ATL-PHYS-PUB-2020-017.

\bibitem{ATLAS:2020lks}
Georges Aad et~al.
\newblock {Search for $ t\overline{t} $ resonances in fully hadronic final
  states in $pp$ collisions at $ \sqrt{s} $ = 13 TeV with the ATLAS detector}.
\newblock {\em JHEP}, 10:061, 2020.

\bibitem{ATLAS:2018wis}
Morad Aaboud et~al.
\newblock {Performance of top-quark and $W$-boson tagging with ATLAS in Run 2
  of the LHC}.
\newblock {\em Eur. Phys. J. C}, 79(5):375, 2019.

\bibitem{CMS:2017ucf}
Albert~M Sirunyan et~al.
\newblock {Search for $ \mathrm{t}\overline{\mathrm{t}} $ resonances in highly
  boosted lepton+jets and fully hadronic final states in proton-proton
  collisions at $ \sqrt{s}=13 $ TeV}.
\newblock {\em JHEP}, 07:001, 2017.

\bibitem{CMS:2021beq}
Albert~M Sirunyan et~al.
\newblock {Search for top squark production in fully-hadronic final states in
  proton-proton collisions at $\sqrt{s} =$ 13 TeV}.
\newblock {\em Phys. Rev. D}, 104(5):052001, 2021.

\bibitem{Sahu:2023uwb}
Rameswar Sahu and Kirtiman Ghosh.
\newblock {ML-Based Top Taggers: Performance, Uncertainty and Impact of Tower
  \& Tracker Data Integration}.
\newblock 9 2023.

\bibitem{Kao:2011aa}
Chung Kao, Hai-Yang Cheng, Wei-Shu Hou, and Joshua Sayre.
\newblock {Top Decays with Flavor Changing Neutral Higgs Interactions at the
  LHC}.
\newblock {\em Phys. Lett. B}, 716:225--230, 2012.

\bibitem{Buchkremer:2013bha}
Mathieu Buchkremer, Giacomo Cacciapaglia, Aldo Deandrea, and Luca Panizzi.
\newblock {Model Independent Framework for Searches of Top Partners}.
\newblock {\em Nucl. Phys. B}, 876:376--417, 2013.

\bibitem{Degrande:2011rt}
Celine Degrande, Jean-Marc Gerard, Christophe Grojean, Fabio Maltoni, and
  Geraldine Servant.
\newblock {An Effective approach to same sign top pair production at the LHC
  and the forward-backward asymmetry at the Tevatron}.
\newblock {\em Phys. Lett. B}, 703:306--309, 2011.

\bibitem{feynrules1}
Neil~D. Christensen and Claude Duhr.
\newblock {FeynRules - Feynman rules made easy}.
\newblock {\em Comput. Phys. Commun.}, 180:1614--1641, 2009.

\bibitem{feynrules2}
Adam Alloul, Neil~D. Christensen, C\'eline Degrande, Claude Duhr, and Benjamin
  Fuks.
\newblock {FeynRules 2.0 - A complete toolbox for tree-level phenomenology}.
\newblock {\em Comput. Phys. Commun.}, 185:2250--2300, 2014.

\bibitem{Dokshitzer:1997in}
Yuri~L. Dokshitzer, G.~D. Leder, S.~Moretti, and B.~R. Webber.
\newblock {Better jet clustering algorithms}.
\newblock {\em JHEP}, 08:001, 1997.

\bibitem{NSubjettiness}
Jesse Thaler and Ken Van~Tilburg.
\newblock {Identifying Boosted Objects with N-subjettiness}.
\newblock {\em JHEP}, 03:015, 2011.

\bibitem{ATLAS:2022ygn}
Georges Aad et~al.
\newblock {Constraints on spin-0 dark matter mediators and invisible Higgs
  decays using ATLAS 13 TeV $pp$ collision data with two top quarks and missing
  transverse momentum in the final state}.
\newblock {\em Eur. Phys. J. C}, 83(6):503, 2023.

\bibitem{Plehn:HepToptagger}
Tilman Plehn, Michael Spannowsky, Michihisa Takeuchi, and Dirk Zerwas.
\newblock {Stop Reconstruction with Tagged Tops}.
\newblock {\em JHEP}, 10:078, 2010.

\bibitem{scikit-learn}
F.~Pedregosa, G.~Varoquaux, A.~Gramfort, V.~Michel, B.~Thirion, O.~Grisel,
  M.~Blondel, P.~Prettenhofer, R.~Weiss, V.~Dubourg, J.~Vanderplas, A.~Passos,
  D.~Cournapeau, M.~Brucher, M.~Perrot, and E.~Duchesnay.
\newblock Scikit-learn: Machine learning in {P}ython.
\newblock {\em Journal of Machine Learning Research}, 12:2825--2830, 2011.

\bibitem{HLLHC2}
I.~Zurbano~Fernandez et~al.
\newblock {High-Luminosity Large Hadron Collider (HL-LHC): Technical design
  report}.
\newblock 10/2020, 12 2020.

\bibitem{PDG}
R.~L. Workman et~al.
\newblock {Review of Particle Physics}.
\newblock {\em PTEP}, 2022:083C01, 2022.

\bibitem{chen2016xgboost}
Tianqi Chen and Carlos Guestrin.
\newblock Xgboost: A scalable tree boosting system.
\newblock In {\em Proceedings of the 22nd acm sigkdd international conference
  on knowledge discovery and data mining}, pages 785--794, 2016.

\bibitem{schapire2013explaining}
Robert~E Schapire.
\newblock {\em Explaining adaboost}, pages 37--52.
\newblock Springer Berlin Heidelberg, Berlin, Heidelberg, 2013.

\bibitem{Cun:DNN}
Yann LeCun, Yoshua Bengio, and Geoffrey Hinton.
\newblock Deep learning.
\newblock {\em nature}, 521(7553):436, 2015.

\bibitem{schmidhuber2015deep}
J{\"u}rgen Schmidhuber.
\newblock Deep learning in neural networks: An overview.
\newblock {\em Neural networks}, 61:85--117, 2015.

\bibitem{LReLU}
Andrew~L. Maas.
\newblock Rectifier nonlinearities improve neural network acoustic models.
\newblock 2013.

\bibitem{chollet2015keras}
Francois Chollet et~al.
\newblock Keras, 2015.

\bibitem{Gong:2022lye}
Shiqi Gong, Qi~Meng, Jue Zhang, Huilin Qu, Congqiao Li, Sitian Qian, Weitao Du,
  Zhi-Ming Ma, and Tie-Yan Liu.
\newblock {An efficient Lorentz equivariant graph neural network for jet
  tagging}.
\newblock {\em JHEP}, 07:030, 2022.

\bibitem{Sahu:2024sts}
Rameswar Sahu.
\newblock {CapsLorentzNet: Integrating Physics Inspired Features with Graph
  Convolution}.
\newblock 3 2024.

\bibitem{Lin:2018cin}
Joshua Lin, Marat Freytsis, Ian Moult, and Benjamin Nachman.
\newblock {Boosting $H\to b\bar b$ with Machine Learning}.
\newblock {\em JHEP}, 10:101, 2018.

\bibitem{10095095}
Bowen Tan, Linfeng Xu, Zihuan Qiu, Qingbo Wu, and Fanman Meng.
\newblock Mfat: A multi-level feature aggregated transformer for person
  re-identification.
\newblock In {\em ICASSP 2023 - 2023 IEEE International Conference on
  Acoustics, Speech and Signal Processing (ICASSP)}, pages 1--5, 2023.

\bibitem{Cowan:2010js}
Glen Cowan, Kyle Cranmer, Eilam Gross, and Ofer Vitells.
\newblock {Asymptotic formulae for likelihood-based tests of new physics}.
\newblock {\em Eur. Phys. J. C}, 71:1554, 2011.
\newblock [Erratum: Eur.Phys.J.C 73, 2501 (2013)].

\bibitem{Gutierrez:2020eby}
Phillip Gutierrez, Rishabh Jain, and Chung Kao.
\newblock {Flavor changing top decays to charm and a Higgs boson with
  $\tau\tau$ at the LHC}.
\newblock {\em Phys. Rev. D}, 103(11):115020, 2021.

\bibitem{Barbier:2004ez}
R.~Barbier et~al.
\newblock {R-parity violating supersymmetry}.
\newblock {\em Phys. Rept.}, 420:1--202, 2005.

\bibitem{Cetinkaya:2020yjfvlq}
V.~Cetinkaya, A.~Ozansoy, V.~Ari, O.~M. Ozsimsek, and O.~Cakir.
\newblock {Single production of vectorlike Y quarks at the HL-LHC}.
\newblock {\em Nucl. Phys. B}, 973:115580, 2021.

\bibitem{ShapBook}
A.E. Roth.
\newblock {\em The Shapley value: essays in honor of Lloyd S. Shapley}.
\newblock Cambridge University Press, 1988.

\bibitem{2022arXiv220205594R}
Benedek {Rozemberczki}, Lauren {Watson}, P{\'e}ter {Bayer}, Hao-Tsung {Yang},
  Oliv{\'e}r {Kiss}, Sebastian {Nilsson}, and Rik {Sarkar}.
\newblock {The Shapley Value in Machine Learning}.
\newblock {\em arXiv e-prints}, page arXiv:2202.05594, February 2022.

\bibitem{land.lee}
Su-In~Lee. Lundberg, Scott~M.
\newblock A unified approach to interpreting model predictions.
\newblock {\em Advances in Neural Information Processing Systems}, 2017.

\bibitem{violplt}
Jerry~L. Hintze and Ray~D. Nelson.
\newblock {Violin Plots: A Box Plot-Density Trace Synergism}.
\newblock {\em The American Statistician}, 52(2):181--184, 1998.

\bibitem{Alwall:2011uj}
Johan Alwall, Michel Herquet, Fabio Maltoni, Olivier Mattelaer, and Tim
  Stelzer.
\newblock {MadGraph 5 : Going Beyond}.
\newblock {\em JHEP}, 06:128, 2011.

\bibitem{NNPDF:2014otw}
Richard~D. Ball et~al.
\newblock {Parton distributions for the LHC Run II}.
\newblock {\em JHEP}, 04:040, 2015.

\bibitem{Bierlich:2022pfr}
Christian Bierlich et~al.
\newblock {A comprehensive guide to the physics and usage of PYTHIA 8.3}.
\newblock 3 2022.

\bibitem{hepmc}
Matt Dobbs and Jorgen~Beck Hansen.
\newblock {The HepMC C++ Monte Carlo event record for High Energy Physics}.
\newblock {\em Comput. Phys. Commun.}, 134:41--46, 2001.

\bibitem{delphes}
J.~de~Favereau, C.~Delaere, P.~Demin, A.~Giammanco, V.~Lema\^\i{}tre,
  A.~Mertens, and M.~Selvaggi.
\newblock {DELPHES 3, A modular framework for fast simulation of a generic
  collider experiment}.
\newblock {\em JHEP}, 02:057, 2014.

\bibitem{fastjet1}
Matteo Cacciari, Gavin~P. Salam, and Gregory Soyez.
\newblock {FastJet User Manual}.
\newblock {\em Eur. Phys. J. C}, 72:1896, 2012.

\bibitem{fastjet2}
Matteo Cacciari and Gavin~P. Salam.
\newblock {Dispelling the $N^{3}$ myth for the $k_t$ jet-finder}.
\newblock {\em Phys. Lett. B}, 641:57--61, 2006.

\bibitem{antikt}
Matteo Cacciari, Gavin~P. Salam, and Gregory Soyez.
\newblock {The anti-$k_t$ jet clustering algorithm}.
\newblock {\em JHEP}, 04:063, 2008.

\bibitem{Krohn:2009zg}
David Krohn, Jesse Thaler, and Lian-Tao Wang.
\newblock {Jets with Variable R}.
\newblock {\em JHEP}, 06:059, 2009.

\bibitem{Mukhopadhyaya:2023rsb}
Biswarup Mukhopadhyaya, Tousik Samui, and Ritesh~K. Singh.
\newblock {Dynamic radius jet clustering algorithm}.
\newblock {\em JHEP}, 04:019, 2023.

\bibitem{Krohn:2009th}
David Krohn, Jesse Thaler, and Lian-Tao Wang.
\newblock {Jet Trimming}.
\newblock {\em JHEP}, 02:084, 2010.

\bibitem{ATLAS:2017bcq}
{Optimisation and performance studies of the ATLAS $b$-tagging algorithms for
  the 2017-18 LHC run}.
\newblock 2017.
\newblock All figures including auxiliary figures are available at
  https://atlas.web.cern.ch/Atlas/GROUPS/PHYSICS/PUBNOTES/ATL-PHYS-PUB-2017-013.

\bibitem{ATLAS:2015prs}
{Performance and Calibration of the JetFitterCharm Algorithm for c-Jet
  Identification}.
\newblock 1 2015.

\bibitem{CMS:2018tuo}
Albert~M Sirunyan et~al.
\newblock {Search for long-lived particles with displaced vertices in multijet
  events in proton-proton collisions at $\sqrt{s}= $13 TeV}.
\newblock {\em Phys. Rev. D}, 98(9):092011, 2018.

\bibitem{ATLAS:2019fwx}
Georges Aad et~al.
\newblock {Search for displaced vertices of oppositely charged leptons from
  decays of long-lived particles in $pp$ collisions at $\sqrt {s}$ =13 TeV with
  the ATLAS detector}.
\newblock {\em Phys. Lett. B}, 801:135114, 2020.

\bibitem{ATLAS:2019kpx}
Georges Aad et~al.
\newblock {Search for heavy neutral leptons in decays of $W$ bosons produced in
  13 TeV $pp$ collisions using prompt and displaced signatures with the ATLAS
  detector}.
\newblock {\em JHEP}, 10:265, 2019.

\bibitem{Chakraborty:2020cpa}
Amit Chakraborty, Stefano Moretti, Claire~H. Shepherd-Themistocleous, and Harri
  Waltari.
\newblock {Extraction of neutrino Yukawa parameters from displaced vertices of
  sneutrinos}.
\newblock {\em JHEP}, 06:027, 2021.

\end{thebibliography}

\end{document}